\documentclass[12pt,a4paper,onecolumn]{article}
\usepackage{graphicx}
\usepackage{amssymb}
\begin{document}
\input epsf
\epsfverbosetrue
\setcounter{page}{1}
\pagenumbering{arabic}
\vspace{66pt}
\begin{center}

\vspace*{15mm}
{\bf 
LAQGSM03.03 Upgrade and its Validation}\\
\vspace{33pt}
{\bf
S. G. Mashnik$^{1}$,
K. K. Gudima$^{2}$,
N. V. Mokhov$^{3}$,
R. E. Prael$^{1}$,
}

\vspace{-1mm}
$^1$X-3-MCC, Los Alamos National Laboratory, Los Alamos, New Mexico 87545, USA\\
$^{2}$Institute of Applied Physics, Academy of Science of
Moldova, Chi\c{s}in\u{a}u, Moldova\\
$^{3}$Fermi National Accelerator Laboratory, MS 220, Batavia, Illinois 
60510-0500, USA

\end{center}
\begin{center}
{\bf Abstract}
\end{center}
This paper presents part of an internal LANL Progress Report
on
LAQGSM03.03, an upgrade of the Los Alamos version of the
Quark-Gluon String Model event generator for MCNPX/6 and MARS15 
transport codes and on its validation and testing against a 
large variety of recent measurements.
We present here an analysis with LAQGSM03.03 of the recent PHENIX
mid-rapidity spectra of $\pi^+$, $\pi^-$, $K^+$, $K^-$, $p$, and $\bar{p}$
produced in ultra-relativistic $p + p$ interactions at $\sqrt{s} = 200$ GeV;
GSI cross sections for the fragmentation of $^{208}$Pb at 1 GeV/nucleon
on $^9$Be; fragmentation cross sections of $^{28}$Si
on  H, C, Al, Cu, Sn, and Pb at energies from 290 to 1200
MeV/nucleon measred recently at HIMAC and BNL; recent HIMAC data
on B, Be, Li, and He
production cross sections from fragmentation of 
$^{12}$C on  H, C, Al, Cu, Sn, and Pb at 290 and 400 MeV/nucleon;
BNL data on fragmentation cross
sections of $^{56}$Fe on  H, C, Al, Cu, and Pb targets at 1.05 GeV/nucleon;
recent  $\pi^+$ and $\pi^-$ spectra 
from 6.4, 12.3, and 17.5 GeV/c p + $^9$Be
from the  E910 BNL measurements;
and fragmentation cross sections
of $^{40}$Ca, $^{48}$Ca, $^{58}$Ni, and  $^{64}$Ni
on  $^9$Be and $^{181}$Ta 
at 140 MeV/nucleon, and of  $^{86}$Kr at 64 MeV/nucleon on the same 
targets measured recently at NSCL-MSU and RARF-RIKEN,
respectively.

{\bf 1. Introduction}

During recent years, for a number of applications like 
Accelerator Transmutation of nuclear Wastes (ATW), 
Accelerator Production of Tritium (APT), 
Spallation Neutron Source (SNS), 
Rare Isotope Accelerator (RIA), 
Proton Radiography (PRAD) as a radiographic probe for the Advanced 
Hydro-test Facility, astrophysical work for NASA, and other projects,
we have developed at the Los Alamos National Laboratory
improved versions\cite{PhotoCEM,CEM03.01} of the Cascade-Exciton Model 
(CEM) \cite{CEM}, to describe  nucleon-, pion-, 
and photon-induced reactions at incident energies up to 
about 5 GeV 
and the Los Alamos version of the Quark-Gluon String Model (LAQGSM) 
\cite{LAQGSM,Varenna06},
to describe reactions induced by particles and
nuclei at energies up to about 1 TeV/nucleon 
(see further references in \cite{Pavia05}--\cite{MCNP6VandV}.

We present here the latest version of LAQGSM, LAQGSM03.03,
which in comparison with its predecessors, 
is developed to describe better 
nuclear reactions at very high energies (above 20 GeV/nucleon),
and which uses, for consistency,
the preequilibrium, evaporation, fission, and Fermi break-up models
in exactly the same form as developed previously 
for the latest version
of our low-energy event generator CEM03.02 \cite{MCNP6VandV};
no longer produces the light unstable final products 
$^6$B, 
$^6$Be, $^5$Li
$^6$H, or $^5$H, that could be produced in
the previous versions of LAQGSM 
in very rare cases via some very asymmetric fission events 
(compare the results shown in Tabs.\ 1 and 2 of Ref. \cite{LAQGSM03.03}): 
LAQGSM03.03 causes 
such unstable products to disintegrate via Fermi breakup
independently of their excitation energy. Finally, some bugs and
small errors observed in previous versions of LAQGSM are fixed;
many useful comments are added.

{\bf 2. LAQGSM03.03 Upgrade}

The code LAQGSM03.03 described here is the latest modification of
LAQGSM~\cite{LAQGSM}, which in its turn is an improvement 
of the Quark-Gluon String Model (QGSM) \cite{QGSM}.
It describes reactions induced by both particles and nuclei, 
as a three-stage process: Intra-Nuclear Cascade (INC), followed
by preequilibrium emission of particles during the equilibration of the
excited residual nuclei formed after the INC, followed by evaporation 
of particles from compound nuclei or fission. 
When the cascade stage of a reaction is completed, we use the
coalescence model described in Refs.~\cite{DCM,Gudima:83a}
to ``create" high-energy d, t, $^3$He, and $^4$He by
final state interactions among emitted cascade nucleons, already outside 
of the target and projectile nuclei.
If the excited compound nucleus produced after the preequilibrium
stage of a reaction is heavy enough ($Z \ge 65$), it may fission,
with subsequent evaporation of particles from the fission fragments.
Such processes are described by LAQGSM03.03 using an 
improved and updated version
of the Generalized Evaporation/fission Model (GEM2)
by Furihata \cite{GEM2}.
On the other hand, if the excited nucleus produced after the 
fast INC stage of a reaction, during emission of particles at the 
preequilibrium or evaporation stages of reaction, or if the fission 
fragment produced via a very asymmetric fission 
becomes quite light ($A < 13$)
LAQGSM03.03 describes its further cooling and disintegration using
the Fermi break-up model, based on the seminal ideas of 
Bohr and Fermi \cite{Fermi50}, instead of using
the preequilibrium and evaporation models.
An illustrative scheme of nuclear reaction calculations by LAQGSM03.03
is shown in Fig. 1.


\begin{figure}[ht]                                                 
\centering
\includegraphics[height=140mm,angle=-0]{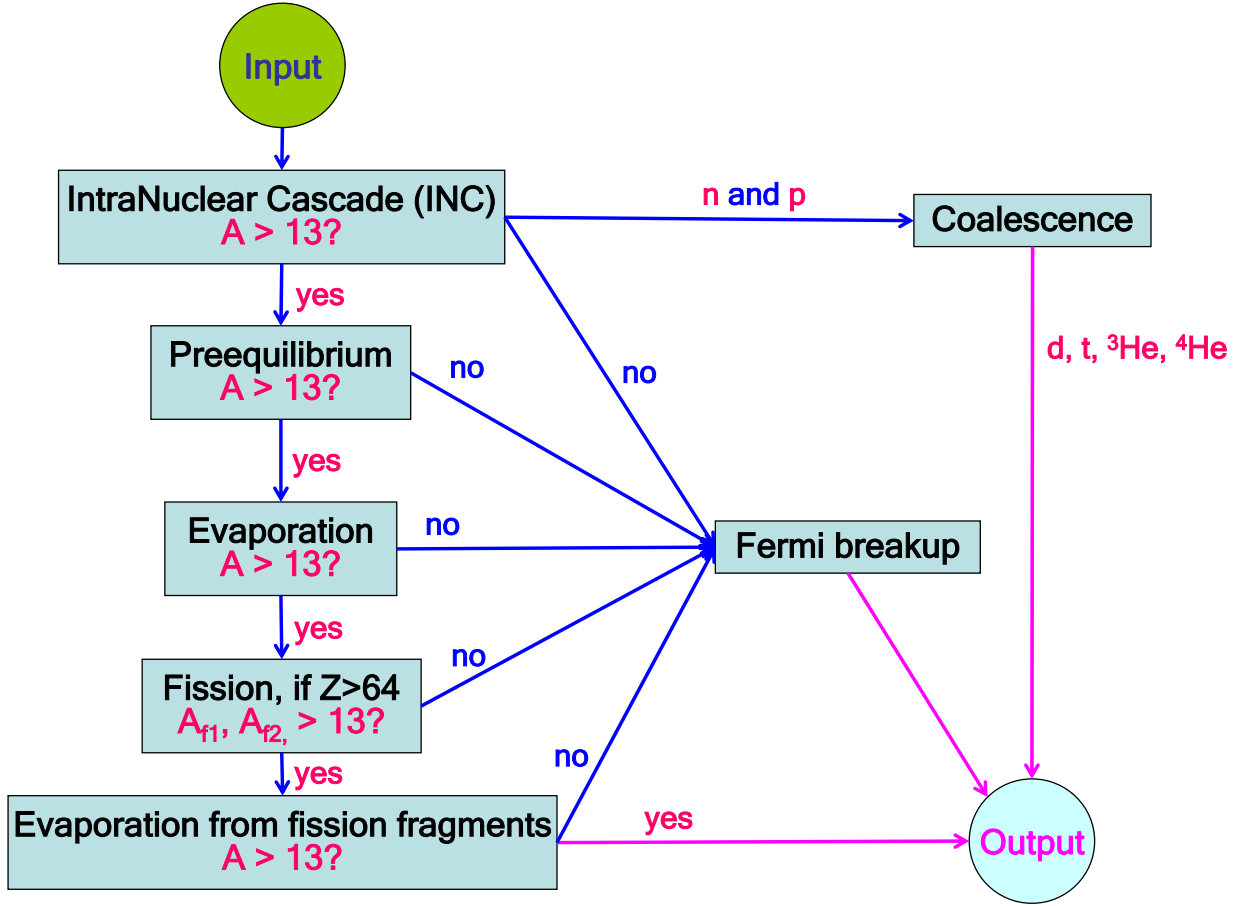}
\caption{General scheme of nuclear reaction calculations by LAQGSM03.03.}
\end{figure}

Striving to make the predictive power of LAQGSM as high as possible,
we have revised, updated, and improved the nuclear reaction 
models used in our event generator.
A brief listing of the physics and of the major
recent improvements in LAQGSM follows.\\

\vspace*{5mm}
{\noindent {\it INC}}

The first and fastest stage of reactions is described by LAQGSM
with a recently improved version \cite{Varenna06,ResNote05}
of the time-dependent intra-nuclear cascade model developed 
initially at JINR in Dubna, often referred to
in the literature as the Dubna intra-nuclear Cascade Model, DCM
(see \cite{DCM}  and references therein). 
The DCM models interactions of fast cascade particles (``participants")
with nucleon spectators of both the target and projectile nuclei and
includes as well interactions of two participants (cascade particles). It
uses experimental cross sections (or those calculated by the Quark-Gluon 
String Model \cite{QGSM,Amelin:86,Kaidalov:87,Amelin:89t}
for energies above 4.5 GeV/nucleon) for these
elementary interactions to simulate angular and energy distributions
of cascade particles, and also considers the Pauli exclusion
principle. In contrast to the earlier versions \cite{EarlyINC,Book} 
of the INC developed at Dubna and utilized with our recent revision 
and improvement in CEM03.01 \cite{CEM03.01}, 
DCM uses a continuous
nuclear density distribution (instead of the approximation of several
concentric zones, where inside each the nuclear density is considered
to be constant); therefore, it
does not need to consider refraction and reflection
of cascade particles inside or on the border of a nucleus;
it also keeps track of the time of an intra-nuclear collision
and of the depletion of the nuclear density during the
development of the cascade (the so-called ``trawling effect").

Recently, we developed \cite{ResNote05}
new approximations to describe more accurately experimental 
elementary energy and angular distributions of secondary particles from 
hadron-hadron and photon-hadron interactions using available data and 
approximations published by other authors. 
The condition for transition from 
the INC stage  of a reaction to preequilibrium was changed; on the whole, 
the INC stage in LAQGSM03.03 is longer while the preequilibrium stage
is shorter in comparison with earlier versions.
A new, high-energy photonuclear reaction model was developed and 
incorporated \cite{Varenna06} into the INC of LAQGSM, 
that allows us to calculate reactions induced by 
photons of up to tens of GeV energy. 
The algorithms of many INC routines
were changed and some INC routines were rewritten, which speeded
up the code significantly; some preexisting bugs in the DCM were fixed;
many useful comments were added.

Specifically for LAQGSM03.03 we have modified our INC for a better 
description of nuclear reactions at very high energies (above 20 GeV/nucleon), 
namely:

1) We have incorporated into LAQGSM the latest fits to currently available
evaluated experimental database for the total and elastic 
$\pi^+ p$, $\pi^- p$, $p p$, and $p n$ 
cross sections (see Chapter 40 in the last Review of Particle Physics 
\cite{PDG06}
and references therein). We use in LAQGSM03.03 these approximations at energies 
above
20--30 GeV, and our own approximations developed for CEM03.01 \cite{CEM03.01}
at lower energies.

2) Previously, we have used LAQGSM only at energies below 800 GeV. 
We studied recently the possibility of 
using LAQGSM03.03 at ultra-relativistic energies, above 1 TeV. 
Our results show that 
to describe ultra-high energy reactions, the 
value of the  parameter  $\sigma_{\perp} = 0.51$ GeV/c in 
the transverse momentum distribution of the constituent quarks of QGSM
(see Eq.\ (12) in \cite{LAQGSM}
or Eq.\ (10) in the first paper of Ref.\ [20])
has to be increased. As shown in Fig.\ 2,
to describe properly $p + p$ interactions at $\sqrt{s} = 200$ GeV,
which corresponds to $T_p \simeq 21314$ GeV, we need to use   
$\sigma_{\perp} = 2.0$ GeV/c. In other words, to be able to describe well with 
LAQGSM
reactions induced by intermediate and high energy projectiles as well as
reactions induced by ultra-relativistic energy projectiles, we need to use 
an energy dependent  average transverse momentum  parameter 
$\sigma_{\perp}$ increasing with the projectile energy from 0.51 GeV/c 
at $T_p \leq 200$ GeV \cite{LAQGSM}
to $\sigma_{\perp} \simeq 2$ GeV/c at  $T_p \simeq 21$ TeV.

\vspace*{5mm}
{\noindent {\it Preequilibrium (PREC)}}

LAQGSM03.03 uses the latest version of the Modified Exciton Model (MEM)
\cite{MEM} as implemented into the latest Cascade-Exciton Model
code CEM03.02 \cite{MCNP6VandV} (and in the publicly available
from RSICC version CEM03.01 \cite{CEM03.01}) 
to describe the relaxation of the nuclear excitation 
of nuclei produced in a reaction after the INC.
MEM takes into account
all possible nuclear transitions changing the number of excitons $n$
with 
$\Delta n = +2, -2$, and 0, and considers all possible multiple subsequent
emissions of $n$, $p$, $d$, $t$, $^3$He, and  $^4$He.
It assumes an equidistant-level scheme with the single-particle
density $g$ and takes into account corrections for the exclusion principle and
indistinguishability of identical excitons.
By neglecting the difference of matrix elements with different $\Delta n$,
$M_+ = M_- = M_0 = M$, MEM estimates the value of $M$ for a given 
nuclear state by associating the $\Delta n = + 2$
transition with the
probability for a quasi-free scattering of a nucleon
above the Fermi level on a nucleon of the target nucleus,
using systematics of available experimental nucleon-nucleon cross sections.

The condition for transition from the preequilibrium stage of a reaction 
to evaporation/fission is changed in comparison with the initial version of 
CEM \cite{CEM}; on the whole, the preequilibrium stage 
in LAQGSM03.03 is shorter while the evaporation stage is longer in comparison 
with earlier versions. The widths for complex-particle emission are 
changed by fitting the probability  $\gamma_\beta$
of several excitons to ``coalesce" 
into a complex particle that may be emitted during the preequilibrium stage
(see details in \cite{CEM03.01,CEM})
to available experimental data on reactions induced by protons and neutrons. 
We have incorporated into CEM03.01
the Kalbach systematics \cite{Kalbach88} to describe angular distributions
of both preequilibrium nucleons and complex particles at incident
energies up to 210 MeV. At higher energies, we use our own
CEM approach (based on Eqs.\ (32,33) of Ref.~\cite{CEM03.01}).
Algorithms of many PREC routines are changed and almost all PREC routines 
are rewritten, which has speeded up the code significantly.
Finally, some bugs are fixed.\\

\clearpage      

\begin{figure}[ht!]                                                 

\vspace*{5mm}
\centering

\includegraphics[height=190mm,angle=-0]{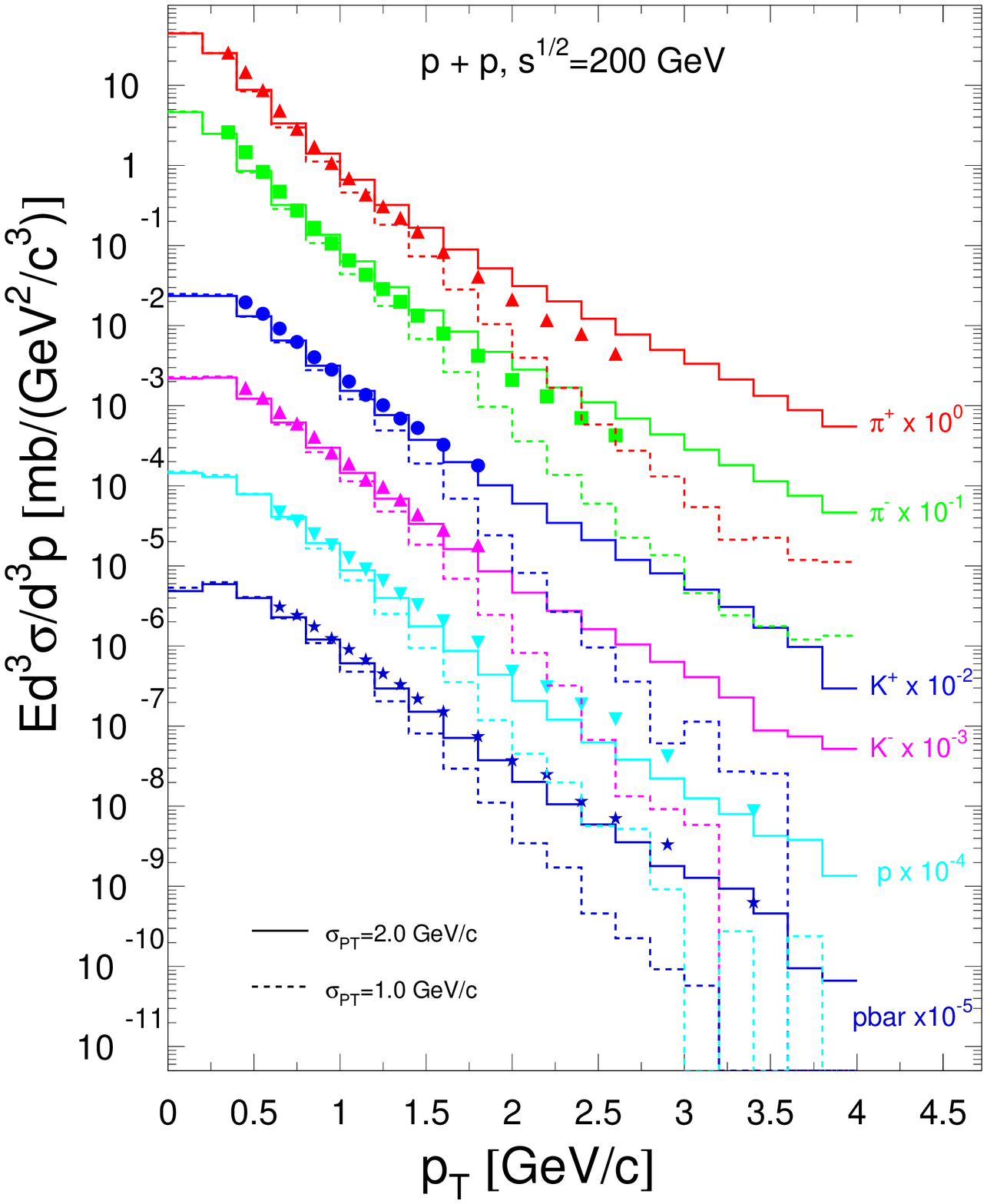}
\caption{}
\end{figure}

Mid-rapidity spectra of $\pi^+$, $\pi^-$, $K^+$, $K^-$, $p$, and $\bar{p}$
produced in ultra-relativistic $p + p$ interactions at $\sqrt{s} = 200$ GeV
($T_p =  21314$ GeV) calculated with
values of the  parameter  $\sigma_{\perp} = 2.0 $ GeV/c (solid histograms)
and  $\sigma_{\perp} = 1.0 $ GeV/c (dashed histograms)
in the transverse momentum distribution of the constituent quarks of the QGSM
compared with recent RHIC data 
\cite{pp200RHIC}

\clearpage

\vspace*{5mm}
{\noindent {\it Evaporation}}

LAQGSM03.03 uses an extension of the Generalized Evaporation Model
(GEM) code GEM2 by Furihata \cite{GEM2}
after the preequilibrium stage of reactions to describe
evaporation of nucleons, complex particles, and 
light fragments heavier than $^4$He (up to $^{28}$Mg)
from excited compound nuclei and to describe their
fission, if the compound nuclei are heavy enough 
to fission ($Z \ge 65$).
GEM describes evaporation with an extension by Furihata
of the Dostrovsky evaporation model~\cite{Dostrovsky},
to include up to 66 types of particles and light fragments that
can be evaporated from an excited compound nucleus.
A very detailed description of GEM2 together with a large amount
of results obtained for many reactions
using GEM2 coupled either with the Bertini INC or with ISABEL
may be found in \cite{GEM2}; many useful details are
presented in \cite{CEM03.01}.\\

\vspace*{5mm}
{\noindent {\it Fission}}

The fission model used in GEM2 is based on the model by Atchison
\cite{RAL},
often referred in the literature as the Rutherford Appleton Laboratory 
(RAL) fission model, which is where Atchison developed it. 
The Atchison fission model was designed to describe only fission of
nuclei with $Z \geq 70$. It assumes that fission competes only with
neutron emission, {\it i.e.}, from the widths $\Gamma_j$ of n, p, d, 
t, $^3$He, and $^4$He emission,
the RAL code calculates the probability of evaporation of these
particles. When a charged particle is selected to be evaporated, 
no fission competition is taken into account. When a neutron is
selected to be evaporated, the code does not actually simulate its
evaporation, instead it considers that fission may compete,
and chooses either fission or evaporation of a neutron according to
the fission probability $P_f$. This quantity is treated by the RAL code 
differently for elements above and below $Z=89$. The mass,
charge, and kinetic-energy distributions of
fission fragments are described by RAL using semi-empirical systematics
developed by Atchison based on experimental data available to him
at that time. 

Furihata used later, more extensive experimental data
and made many changes in the calculation of both the fission widths and
mass, charge, and kinetic-energy distributions of the
fission fragments. Details are given in \cite{CEM03.01,GEM2}.
In comparison with the original
GEM2, the calculation of fission widths in LAQGSM03.03 is changed by
fitting the ratio of the level-density parameters at the saddle point to
those in the evaporation channel to the systematics of proton-induced fission
cross sections by Prokofiev~\cite{Prokofiev}
(see details in \cite{fitaf}). 
This affects as well the relative probabilities
of particle evaporation, in the case of heavy nuclei, where 
competition between evaporation and fission is considered.

In our codes,
we have fixed first several observed uncertainties and small errors
in the 2002 version of GEM2 which Dr.\ Furihata kindly sent us. 
We extend GEM2 to describe fission of lighter nuclei, down to $Z \ge 65$,
and modify it \cite{fitaf}
so that it provides a good description of fission
cross sections when it is used after our INC and preequilibrium models.
Several GEM2 routines are slightly modified in CEM03.01 and
LAQGSM03.03 and some bugs are fixed.\\

\vspace*{5mm}
{\noindent {\it Coalescence}}

The coalescence model implemented in LAQGSM03.03
is described in  Refs.~\cite{DCM,Gudima:83a}.
In contrast to most other
coalescence models for heavy-ion induced reactions,
where complex particle spectra are estimated simply by
convolving the measured or calculated inclusive spectra of nucleons
with corresponding fitted coefficients (see, {\it e.g.}, \cite{Kapusta:80}
and references therein), LAQGSM03.03 uses in its simulations of complex
particle coalescence real information about all emitted cascade nucleons
and does not use convolutions of nucleon spectra. LAQGSM03.03 assumes that
nucleons emitted during the INC stage of a reaction may
form an appropriate composite particle, if they have a
correct isotopic content and the differences in their momenta are
smaller than $p_c$,
equal to 90, 108, and 115 MeV/c for $d$, $t$($^3$He), and $^4$He,
respectively. When, for example, an INC proton coalescences
with an INC neutron into a deuteron, both of them are removed from
the status of nucleons, leaving in the final state only the deuteron.

In comparison with the initial version \cite{DCM,Gudima:83a}, 
in LAQGSM03.03 we have changed/deleted several routines
and have tested them against a large variety
of measured data on nucleon- and
nucleus-induced reactions at different incident energies.\\

\vspace*{5mm}
{\noindent {\it Fermi Breakup}}

The Fermi breakup model \cite{Fermi50}
describes a break-up of an excited nucleus into $n$ components
in the final state ({\it e.g.}, a possible residual nucleus,
nucleons, deuterons, tritons, alphas, {\it etc.}) according
to the $n$-body phase space distribution. The version of the 
Fermi breakup model code used in LAQGSM03.03 was 
developed in the former group of Prof.\ Barashenkov
at the Joint Institute for Nuclear Research (JINR), Dubna, Russia.
The angular distribution of $n$ emitted fragments is assumed to be
isotropic in the c.m.\ system of the disintegrating nucleus and their
kinetic energies are calculated from momentum-energy conservation.
The Monte-Carlo method is used to randomly select the decay channel
according to the corresponding probabilities. 
Then, for a given channel, LAQGSM03.03 calculates kinematic 
quantities for each fragment according to the
$n$-body phase space distribution using the Kopylov method \cite{Kopylov:70}.
Generally, LAQGSM03.03 considers formation of fragments only in their ground 
and those low-lying states which are stable for nucleon emission.
All formulas and algorithms used in the initial version
are described in details by Amelin \cite{GEANT4} and 
may be found in a shorter form in Ref.~\cite{CEM03.01} as well,
therefore we do not repeat them here.

In comparison with its initial versions, we have modified LAQGSM03.03 to 
decay some unstable light fragments that were produced by the original 
Fermi-breakup-model code described in~\cite{GEANT4}. As mentioned above,  
the initial routines that describe the Fermi breakup model were
written more than twenty years ago in the group
of Prof. Barashenkov at JINR, Dubna, and unfortunately had 
some problems.
First, these routines
allowed in rare cases production of some light unstable
fragments like $^5$He, $^5$Li, $^8$Be, $^9$B, {\it etc.}, as a result
of a break-up of some light excited nuclei. Second, they
very rarely allowed even production of
``neutron stars'' (or ``proton stars''), {\it i.e.}, residual ``nuclei''
produced via Fermi breakup that consist of only neutrons (or only protons).
Lastly, these routines could even crash the code, due to cases of 
division by 0.
All these problems of the Fermi breakup model routines were
addressed and solved by Dick Prael for CEM03.02
\cite{MCNP6VandV}; the changes were then put in LAQGSM03.02 \cite{MCNP6VandV}.
Several bugs are also fixed.

However, even after solving these problems and after implementing the 
improved Fermi breakup model into CEM03.02 
and LAQGSM03.02 \cite{MCNP6VandV}, these event generators still could
produce some unstable products via very asymmetric fission, when the
excitation energies of those fragments were below 3 MeV so they were not
checked and disintegrated with the Fermi breakup model. 
Table 1 in Ref. \cite{LAQGSM03.03}
shows an example of such results from an output of the
reaction  1 GeV/nucleon
$^{208}$Pb + $^9$Be calculated with
LAQGSM03.02 \cite{MCNP6VandV}. We can see that 
from a total of 10$^7$ simulated
inelastic events, LAQGSM03.02 produced 60 unstable light fragments, namely:
one $^5$H, twenty-three $^6$H, one $^5$Li, thirty $^6$Be, one $^{13}$Be,
and four $^6$B.
The summed yield 
of all these unstable products is less than 0.0006\% of the 
total yield of all products, so that production of these unstable 
nuclides affects by less than 0.0006\% the other correct
cross sections from this test problem. However, these unstable nuclides are 
non-physical and should be eliminated.
This is the reason
we have incorporated into LAQGSM03.03 a universal checking of all 
unstable light products. We force such unstable products to disintegrate 
via Fermi breakup independently of their excitation energy.
Table 2 of Ref. \cite{LAQGSM03.03}
presents results for the same reaction 
as shown in Tab.\ 1 of that paper, but calculated with LAQGSM03.03.
We can see that
this version does not produce any such unstable light products.

\vspace*{5mm}
{\bf 3. Validation of LAQGSM03.03}

We have tested the LAQGSM03.03 code
against a large variety of particle-particle,
particle-nucleus, and nucleus-nucleus reactions at energies from
$\sim 10$ MeV/nucleon to $\sim 1$ TeV/nucleon,
some measured very recently, and some earlier ones analyzed
already with previous versions of this event generator.
The general agreement of our results with the new experimental data is about
the same as the agreement with the older data analyzed with previous 
versions of LAQGSM and published in 
Refs.~\cite{PhotoCEM}, \cite{LAQGSM}--\cite{MCNP6VandV}, \cite{HSS06}.
Therefore, we present only comparisons of model results compared to 
several very recent measurements.
We note that LAQGSM03.03 is being (or already has been) incorporated 
as the major event generator into
the FNAL MARS15 \cite{MARS} and LANL MCNP6 \cite{MCNP6} and 
MCNPX \cite{MCNPX} transport codes.

Figs.\ 3 and 4 show comparisons of recent GSI measurements \cite{TeresaThesis}
of the fragmentation of $^{208}$Pb on $^9$Be at 1 GeV/nucleon
with results from LAQGSM03.03 and from its previous
version, LAQGSM03.02 \cite{MCNP6VandV} (the same reaction and calculations
as shown in Tabs.\ 1 and 2 of Ref. \cite{LAQGSM03.03} discussed above).
These GSI measurements were done with a special interest in heavy neutron-rich
nuclei approaching the stellar nucleosynthesis $r$-process path around $A = 195$;
they therefore contain experimental data only for products from Yb to Bi, while we
calculate with our codes all possible products and present in Fig.\ 4 
our predictions
for yields of yet unmeasured nuclear products lighter than Yb.
LAQGSM03.03 describes these new GSI data reasonably well and
certainly no worse than its predecessor, also not predicting
unstable non-physical light fragments, as did LAQGSM03.02.

Fig.\ 5 presents part of the recent extensive experimental data
on fragmentation cross
sections of $^{28}$Si on H, C, Al, Cu, Sn, and Pb at energies from 290 to 1200
MeV/nucleon \cite{Zeitlin07}. Such measurements are needed for 
NASA to plan long-duration spaceflights and to test the models

\clearpage      
\begin{figure}[ht]                                                 

\vspace*{-30mm}
\centering

\hspace*{-25mm}
\includegraphics[height=290mm,angle=-0]{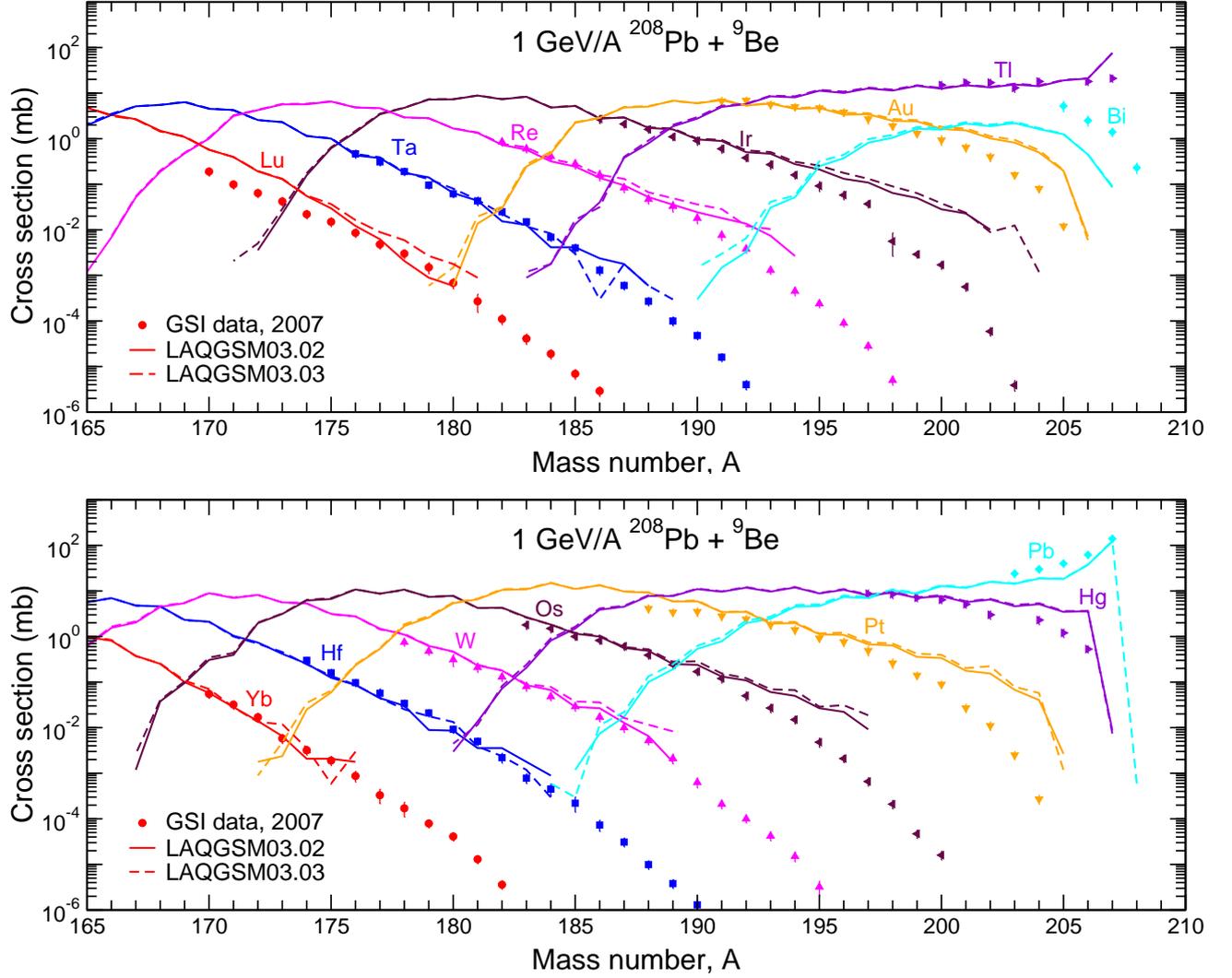}

\vspace*{-110mm}
\caption{Mass-number distribution of the cross section for the production
of thirteen elements from Yb to Bi
from the reaction 1 GeV/nucleon $^{208}$Pb + $^9$Be.
Symbols are GSI measurements of Nieto {\it et al.} [38];
dashed lines are results from LAQGSM03.03, while solid lines are
results from LAQGSM03.02 
\cite{MCNP6VandV}.
}
\end{figure}

\noindent{
used to 
evaluate radiation exposure in flight,
and were performed at many incident energies in this energy range
at the Heavy Ion Medical
Accelerator in Chiba (HIMAC) and at Brookhaven National Laboratory
(see details in \cite{Zeitlin07} and references therein).
}
We calculate in our model practically all these data,
but here limit ourselves to examples of results for only three energies,
for each measured target. For comparison, we present in Fig.\ 5 results from
both LAQGSM03.03 (solid lines) and its predecessor LAQGSM03.02 (dashed lines). 
In general, LAQGSM03.03 describes these new data slightly
better than LAQGSM03.02 \cite{MCNP6VandV}, although this is not obvious
on the scale of the figure.
The agreement of our calculations with these data is excellent, especially 
considering that the results presented in this figure, just as all our other 
results, are obtained without fitting any parameters in the code; we 
simply input $A$ and $Z$ of the projectile and target and the energy of
the projectile, then calculate without changing or fitting anything.

\clearpage      
\begin{figure}[ht]                                                 

\vspace*{-40mm}
\centering
\hspace*{-10mm}
\includegraphics[height=260mm,angle=-0]{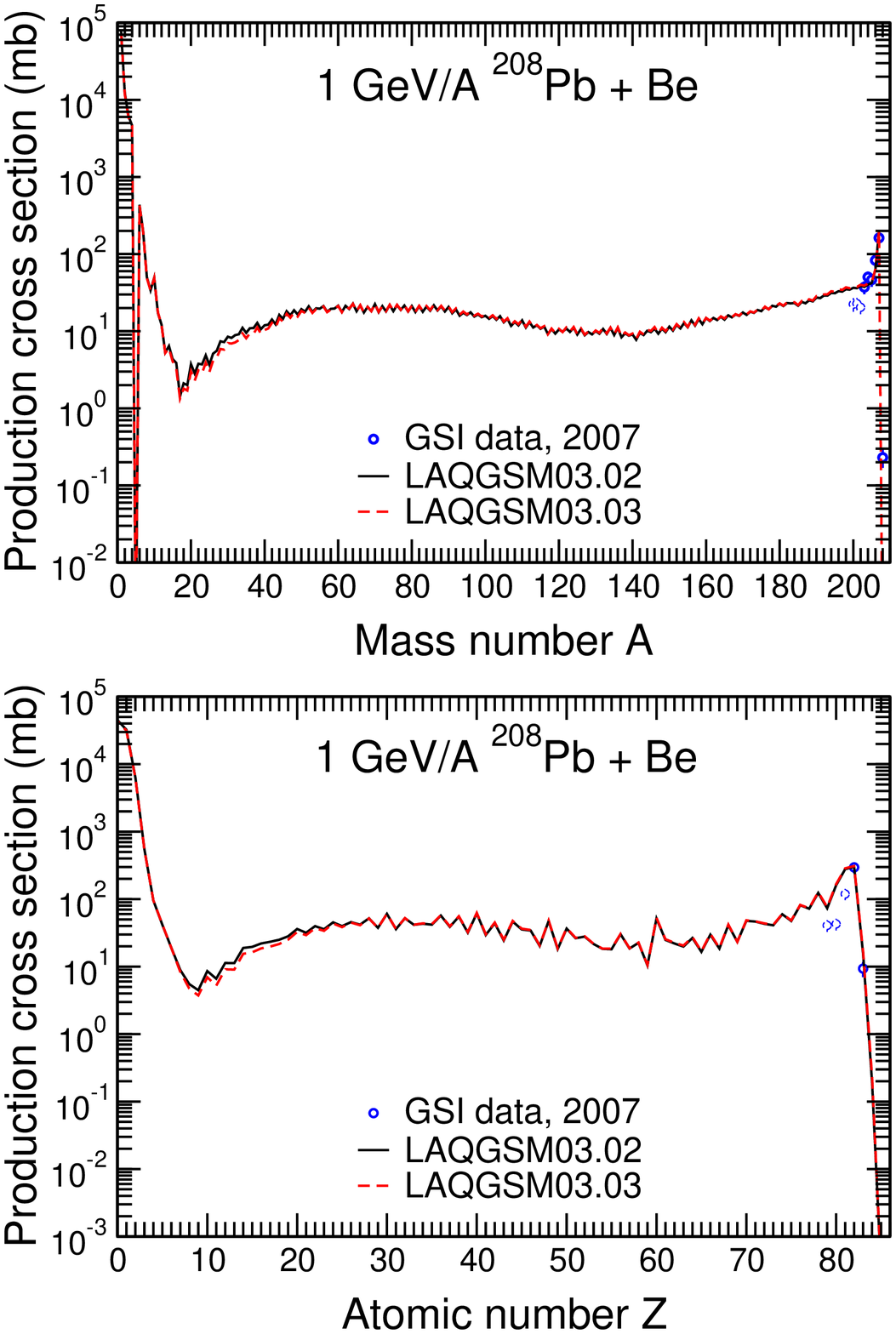}

\vspace*{-15mm}
\caption{Mass- and charge-number distributions of the yield of all
products from the reaction 1 GeV/nucleon $^{208}$Pb + $^9$Be.
Symbols sre GSI measurements of Nieto {\it et al.} 
\cite{TeresaThesis}; 
dashed lines are results from LAQGSM03.03, while solid lines are
results from LAQGSM03.02 
\cite{MCNP6VandV}.
}
\end{figure}

\clearpage      

\begin{figure}[ht]                                                 

\vspace*{-30mm}
\centering

\hspace*{-10mm} 
\includegraphics[height=265mm,angle=-0]{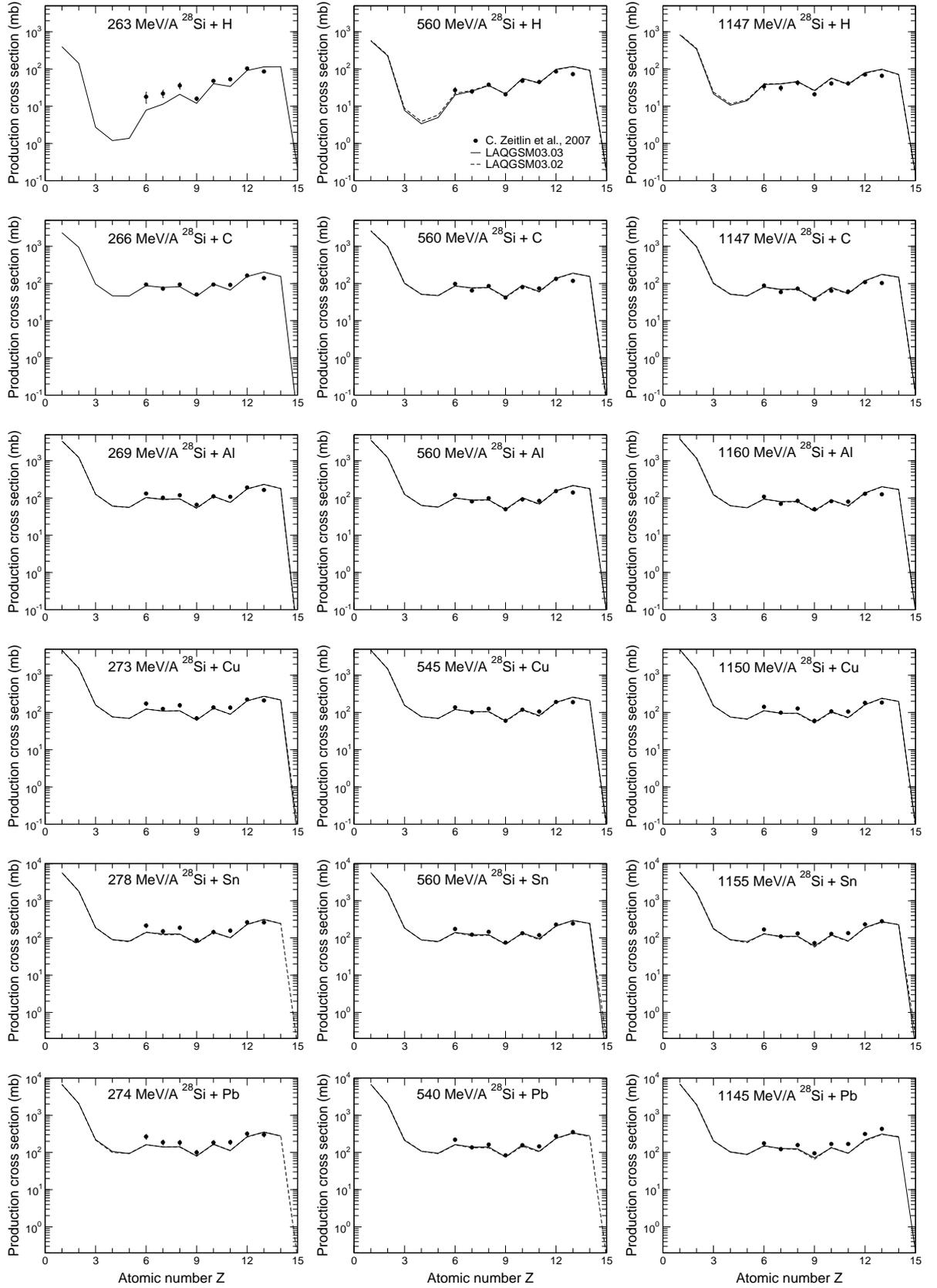}
\vspace*{-30mm}
\caption{Atomic-number dependence of the fragment-production cross sections 
from the interactions of $^{28}$Si of about 270, 560, and 1150 MeV/nucleon
with H, C, Al, Cu, Sn, and Pb, as indicated.
Filled circles are measurements by Zeitlin {\it et al.} 
\cite{Zeitlin07}; 
solid lines are results from LAQGSM03.03, while dashed lines are
results from LAQGSM03.02 
\cite{MCNP6VandV}.
}
\end{figure}

\clearpage            

\begin{figure}[ht]                                                 

\vspace*{-20mm}
\centering

\hspace*{-15mm}
\includegraphics[height=280mm,angle=-0]{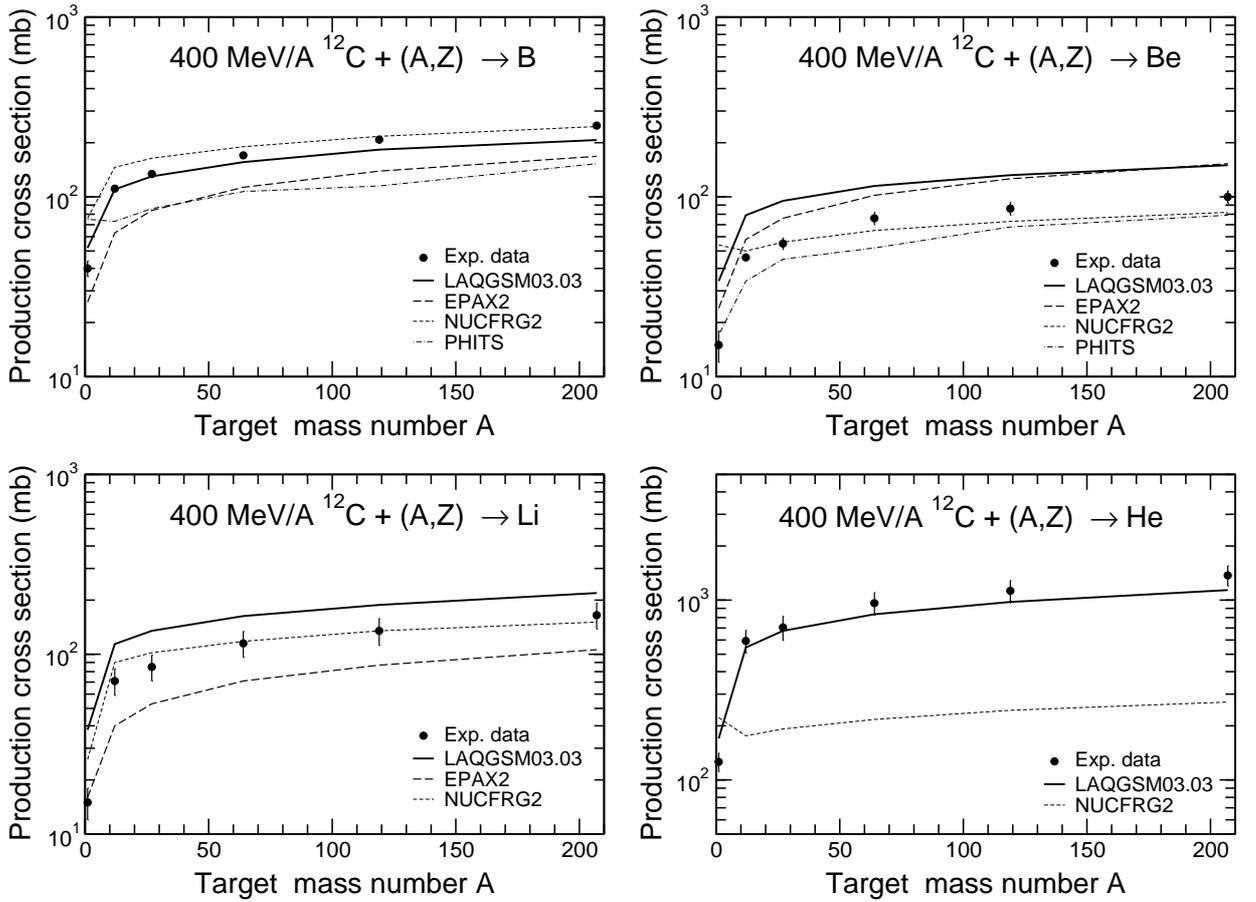}  
\vspace*{-140mm}
\caption{Target mass-number dependence of B, Be, Li, and He
production cross sections from the interactions of 400 MeV/nucleon
$^{12}$C with H, C, Al, Cu, Sn, and Pb.
Filled circles are measurements by Zeitlin {\it et al.} 
\cite{Zeitlin07b}; 
Solid lines are results from LAQGSM03.03 
compared with experimental data and with results
from EPAX2 
\cite{EPAX2},
NUCFRG2 
\cite{NUCFRG2},
and
PHITS 
\cite{PHITS}
taken from Tab.\ VII of Ref.\ 
\cite{Zeitlin07b}.
}
\end{figure}

Fig.\ 6 and 7 show recent data from two more experiments performed at HIMAC
by the same group of Zeitlin {\it et al.}, namely, B, Be, Li, and He yields
from interactions of $^{12}$C with H, C, Al, Cu, Sn, and Pb at
400 and 290 MeV/nucleon, respectively \cite{Zeitlin07b}. These data are of
interest for cancer therapy with carbon ions used currently at several facilities,
as well as for radiation protection of astronauts on long-duration
space missions (see references and details in \cite{Zeitlin07b}).
This is why the authors of the measurements have analyzed their data
with widely used phenomenological systematics EPAX2 \cite{EPAX2},
the one-dimensional 
NASA transport code NUCFRG2 
\cite{NUCFRG2}, and with the recent Japanese transport code PHITS 
\cite{PHITS}; for comparison, results from these codes taken from Tabs.\ IV and VII of 
Ref.~\cite{Zeitlin07b} are also shown in Figs.\ 6 and 7
together with our LAQGSM03.03 results.

\clearpage            

\begin{figure}[ht]                                                 

\vspace*{-20mm}
\centering
\hspace*{-15mm}
\includegraphics[height=280mm,angle=-0]{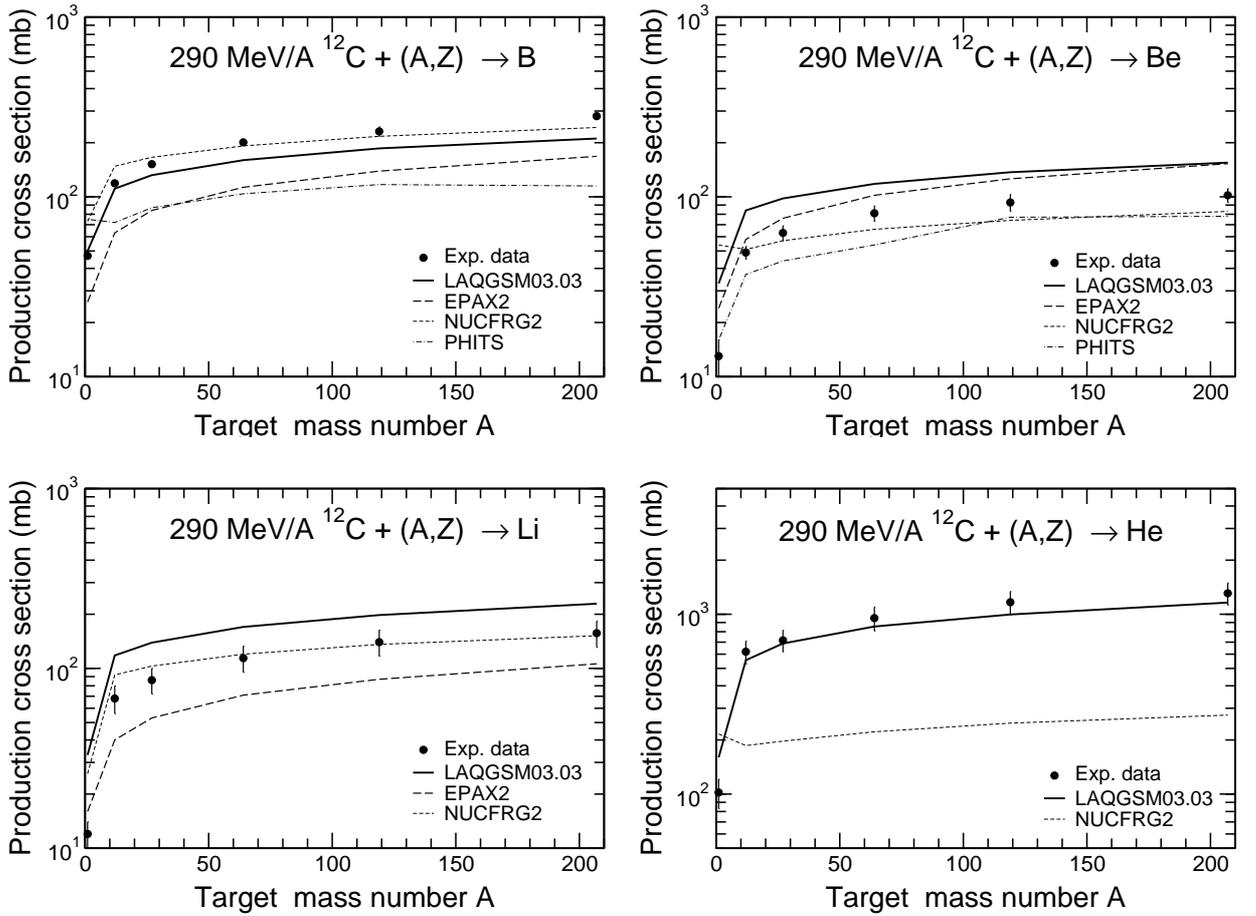}   
\vspace*{-140mm}
\caption{Target mass-number dependence of B, Be, Li, and He
production cross sections from interactions of 290 MeV/nucleon
$^{12}$C with H, C, Al, Cu, Sn, and Pb.
Filled circles are measurements by Zeitlin {\it et al.} 
\cite{Zeitlin07b}. 
Solid lines are results from LAQGSM03.03 
compared with experimental data and with results
from EPAX2 
\cite{EPAX2},
NUCFRG2 
\cite{NUCFRG2},
and
PHITS 
\cite{PHITS}
taken from Tab.\ IV of Ref.\ 
\cite{Zeitlin07b}.
}
\end{figure}

The extracted experimental charge-changing cross sections 
\cite{Zeitlin07b} shown in Figs.\ 6 and 7 were obtained at three
distinct values of angular acceptance, and can not be compared
directly with results of calculations by LAQGSM03.03 or by other
models that do not account for the real complexity of the experiment
(see details in \cite{Zeitlin07b}). However, we see a reasonable 
agreement of our results with these experimental data and with results by
other codes, though a straightforward comparison of calculations
with these data is difficult. On the whole, LAQGSM03.03 agrees
with these measurements no worse than EPAX2 \cite{EPAX2},
NUCFRG2 \cite{NUCFRG2},
and PHITS \cite{PHITS}, and do especially well for for He production.

Fig.\ 8 shows one more set of data measured at Brookhaven National Laboratory
by the same group; namely, fragmentation cross
sections for $^{56}$Fe on  H, C, Al, Cu, and Pb targets at 1.05 GeV/nucleon
\cite{Zeitlin97}, compared with measurements of the same reactions at 
nearby energies of 
1.88 GeV/nucleon by Westfall {\it et al.}
\cite{Westfall79},
1.55 GeV/nucleon by Cummings {\it et al.}
\cite{Cummings90},
and
1.086 GeV/nucleon by Webber {\it et al.}
\cite{Webber90},
as well as with LAQGSM03.03 results.
LAQGSM03.03 describes these data very well.

Fig.\ 9 shows a test of LAQGSM03.03 on
another type of data: inclusive pion
production spectra in proton-beryllium
collisions at 6.4, 12.3, and 17.5 GeV/c obtained from 
data taken by the already quite old E910
measurement at  Brookhaven National Laboratory, but analyzed and published
only a month ago \cite{Chemakin07}. LAQGSM03.03 describes
these pion spectra quite well, just as we obtained with previous
versions of LAQGSM for other
spectra of different ejectiles measured by the E910 experiment.

Finally, Figs.\ 10--19 show a comparison of our results with the
recent extensive measurements by Mocko {\it et al.}
of the projectile fragmentation of 
$^{40}$Ca, $^{48}$Ca, $^{58}$Ni, and $^{64}$Ni at
140 MeV/nucleon on $^9$Be and $^{181}$Ta targets measured at the
National Superconducting Cyclotron Laboratory (NSCL) at
Michigan State University
\cite{Mocko06,MockoPhD}
and of fragmentation of $^{86}$Kr at 64 MeV/nucleon
on the same targets, measured at RIKEN \cite{MockoPhD,Mocko07}.
These measurements are similar in their technique to
experiments done recently at GSI at higher energies,
analyzed with previous versions of LAQGSM
\cite{Pavia05}--\cite{MCNP6VandV}; one example is
shown in Figs.\ 3 and 4. The cross sections for the production 
of different isotopes of different elements obtained in 
this type of measurement
are much more informative and useful for applications,
as well as in developing and testing nuclear-reaction models than are 
the charge-changing integral cross sections 
\cite{Zeitlin07,Zeitlin07b},
\cite{Zeitlin97}--\cite{Webber90} discussed above.
It is much more difficult to describe with a model such detailed
cross sections than to describe integral yields of products,
or spectra of emitted particles; this is why such data are extremely
useful to validate models and codes. If fact, Dr.\ Mocko has analyzed 
\cite{MockoPhD}
these measurements with the empirical parameterization EPAX \cite{EPAX2},
with the more detailed but still semi-phenomenological
Abrasion-Ablation (AA) model \cite{AA}
as implemented into the LISE++ code \cite{LISE} and the
Heavy-Ion Phase-Space Exploration (HIPSE) model \cite{HIPSE},
as well as with the more complicated
Antisymmetrized Molecular Dynamics (AMD) model \cite{AMD}.
Dr.\ Mocko has found \cite{MockoPhD} that
none of these models in their standard versions are able to describe well
the whole set of data \cite{Mocko06,Mocko07}, and all of them would 
need to be improved to 
agree with these measurements. Figs.\ 10--19 show that
LAQGSM03.03 agrees quite well with the whole set of measured cross
sections, especially considering that these calculations are done with
a fixed model, without changing or fitting anything. In fact, these
calculations were done before having numerical values of the experimental 
data. We received from Dr.\ Mocko a list of reactions to
be calculated, performed our calculations and sent 
him the results. He then compared our results with the measurements 
and plotted Figs.\ 10 to 19 (as well as others, to be published in a 
future common paper on this analysis). 
From Figs.\ 10 to 19 we see that the agreement of our 
results with the data \cite{Mocko06,Mocko07} is good but not perfect,
there is room for future improvements of LAQGSM.
But even in its current ``03.03'' version, LAQGSM describes the 
data better than do any other models or phenomenological parameterizations 
so far considered (see details in \cite{MockoPhD}).

\begin{center}
{\it Acknowledgment}
\end{center}

\noindent
We are grateful to Drs.\ Arnold Sierk, Mircea Baznat,
Michal Mocko, and Paolo Napolitani for useful discussions.
This work was carried out under the auspices of the
U.\ S.\ Department of Energy (DOE)
and National Aeronautics and Space Administration (NASA).

\vspace*{-10mm}

\clearpage            

\begin{figure}[ht]                                                 

\vspace*{-20mm}
\centering
\hspace*{-20mm}
\includegraphics[height=290mm,angle=-0]{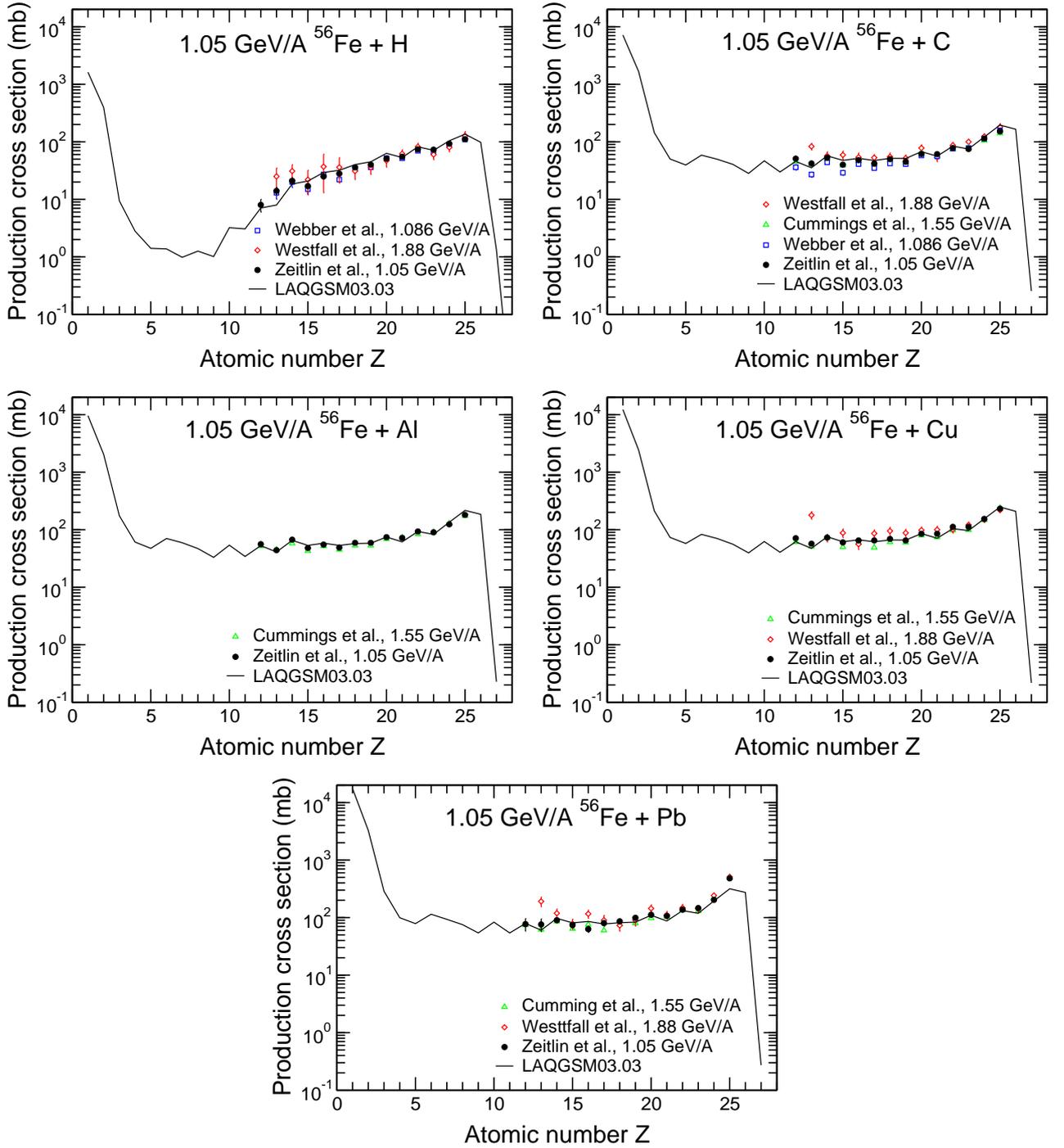}   
\vspace*{-80mm}
\caption{Atomic-number dependence of the fragment production cross sections 
from interactions of 1.05 GeV/nucleon $^{56}$Fe
with H, C, Al, Cu, and Pb.
Filled circles show the 
measurements by Zeitlin {\it et al.} 
\cite{Zeitlin97};
solid lines are results from LAQGSM03.03.
For comparison, measurements of the same reactions at 
nearby energies of 
1.88 GeV/nucleon by Westfall {\it et al.} 
\cite{Westfall79},
1.55 GeV/nucleon by Cummings {\it et al.} 
\cite{Cummings90},
and
1.086 GeV/nucleon by Webber {\it et al.} 
\cite{Webber90},
are shown with colored diamonds, triangles, and squares, respectively.
}
\end{figure}
\clearpage            

\begin{figure}[ht]                                                 

\vspace*{-30mm}
\centering
\hspace*{-13mm}
\includegraphics[height=270mm,angle=-0]{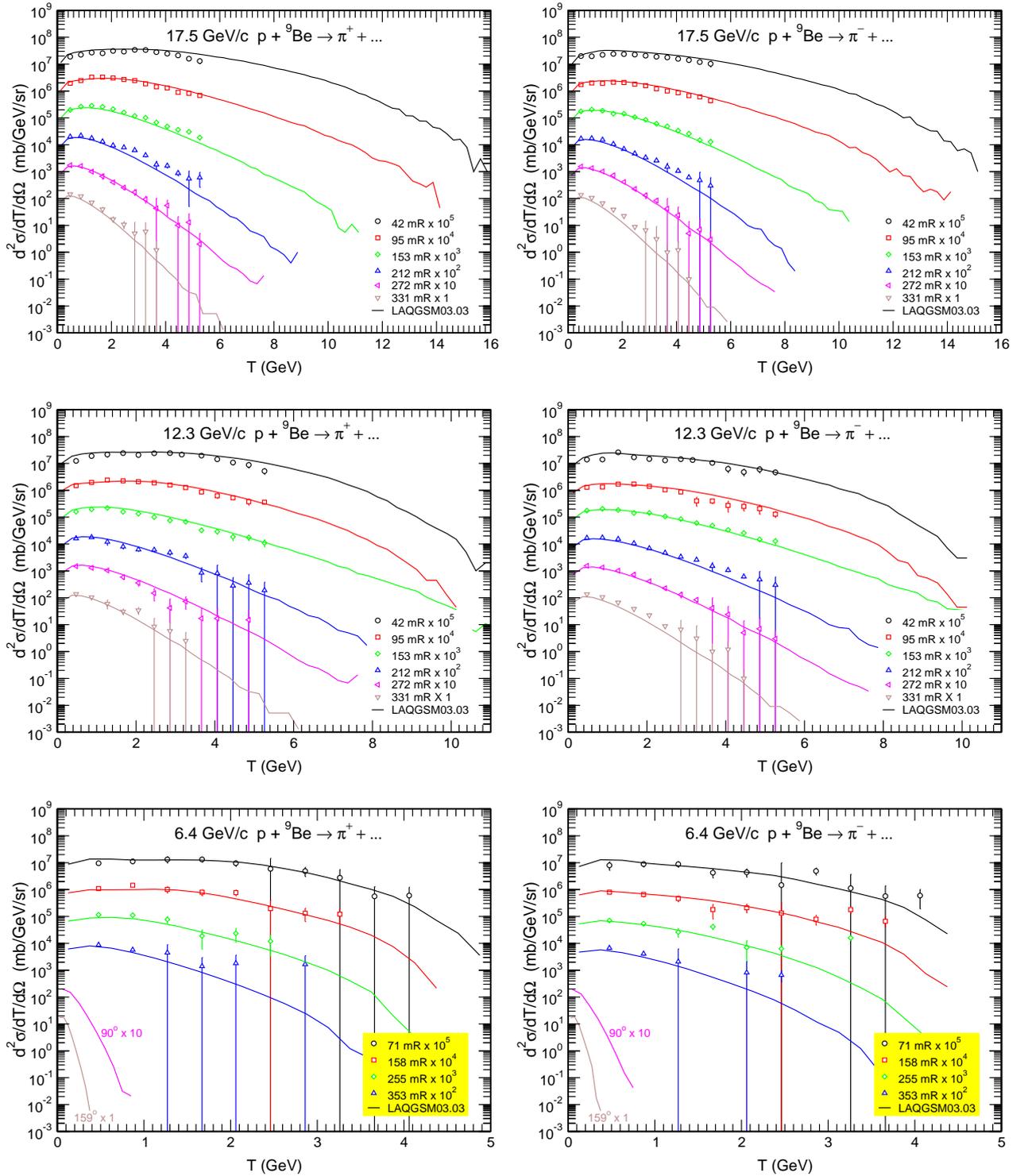} 
\vspace*{-50mm}
\caption{Measured inclusive forward $\pi^+$ and $\pi^-$ spectra 
from 6.4, 12.3, and 17.5 GeV/c p + $^9$Be 
\cite{Chemakin07}
compared with LAQGSM03.03 results at angles 
of detection as indicated in the plots. For reactions induced by 6.4 GeV/c
protons,
we also show LAQGSM03.03 predictions for unmeasured spectra 
at 90 and 159 degrees. 
}
\end{figure}
\clearpage            

\begin{figure}[ht]                                                 

\centering
\hspace*{-5mm}
\includegraphics[height=170mm,angle=-0]{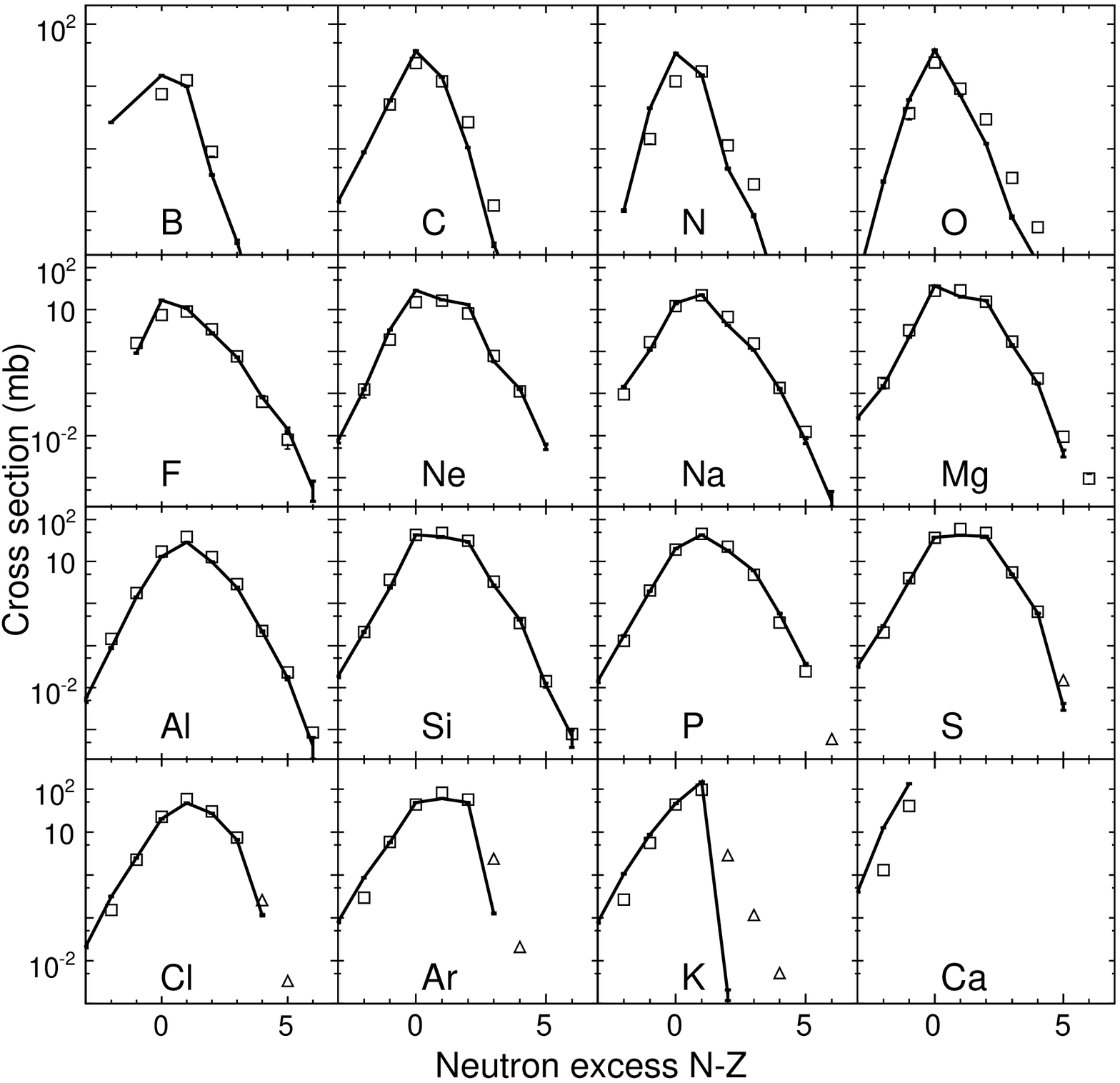} 
\caption{Measured cross sections for $^{40}$Ca fragmentation on $^9$Be
at 140 MeV/nucleon 
\cite{Mocko06,MockoPhD}
compared with LAQGSM03.03 predictions. 
}
\end{figure}
\clearpage            

\begin{figure}[ht]                                                 

\centering
\hspace*{-5mm}
\includegraphics[height=170mm,angle=-0]{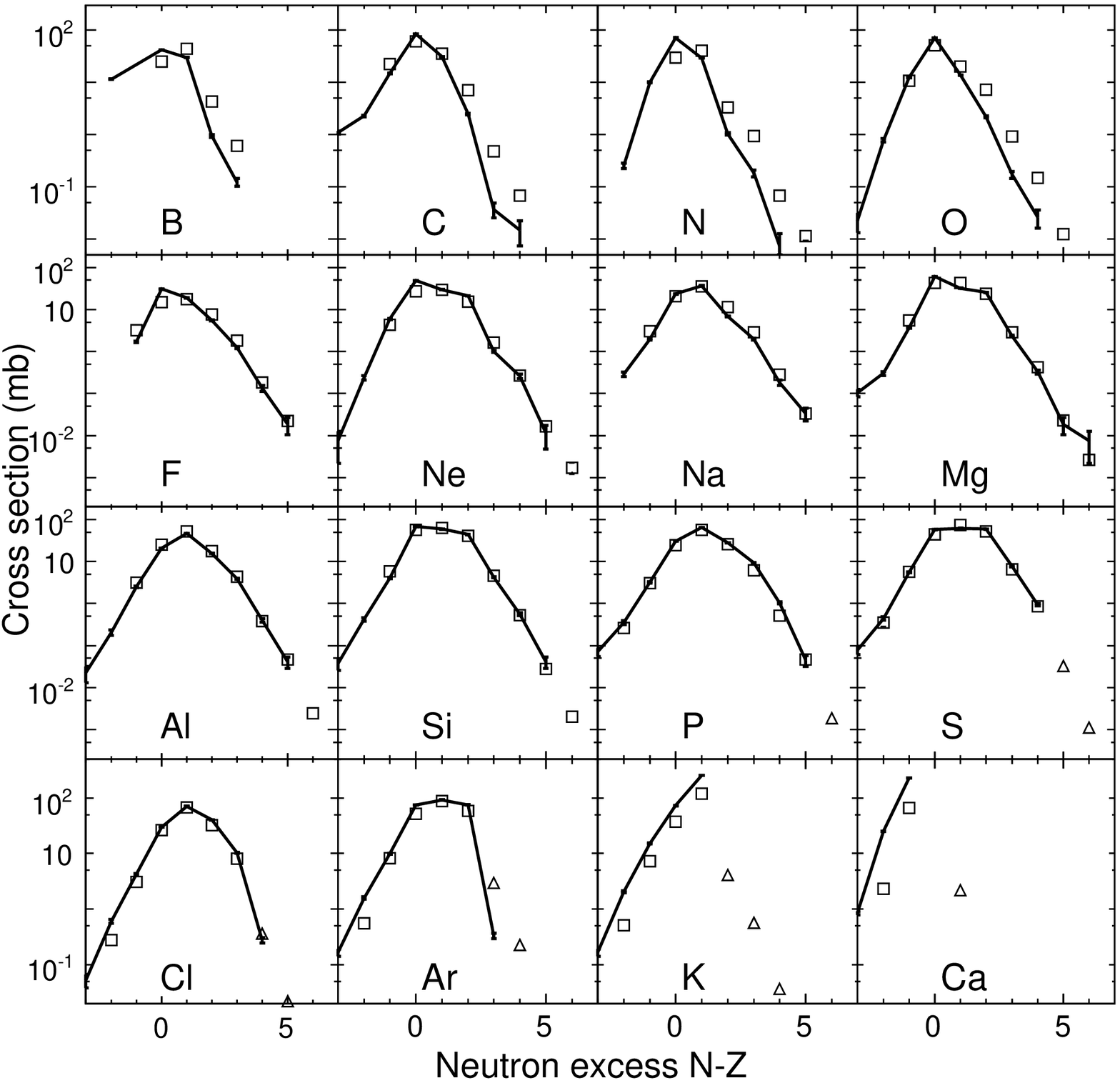} 
\caption{Measured cross sections for $^{40}$Ca fragmentation on $^{181}$Ta
at 140 MeV/nucleon 
\cite{Mocko06,MockoPhD}
compared with LAQGSM03.03 predictions. 
}
\end{figure}
\clearpage            

\begin{figure}[ht]                                                 

\centering
\hspace*{-5mm}
\includegraphics[height=170mm,angle=-0]{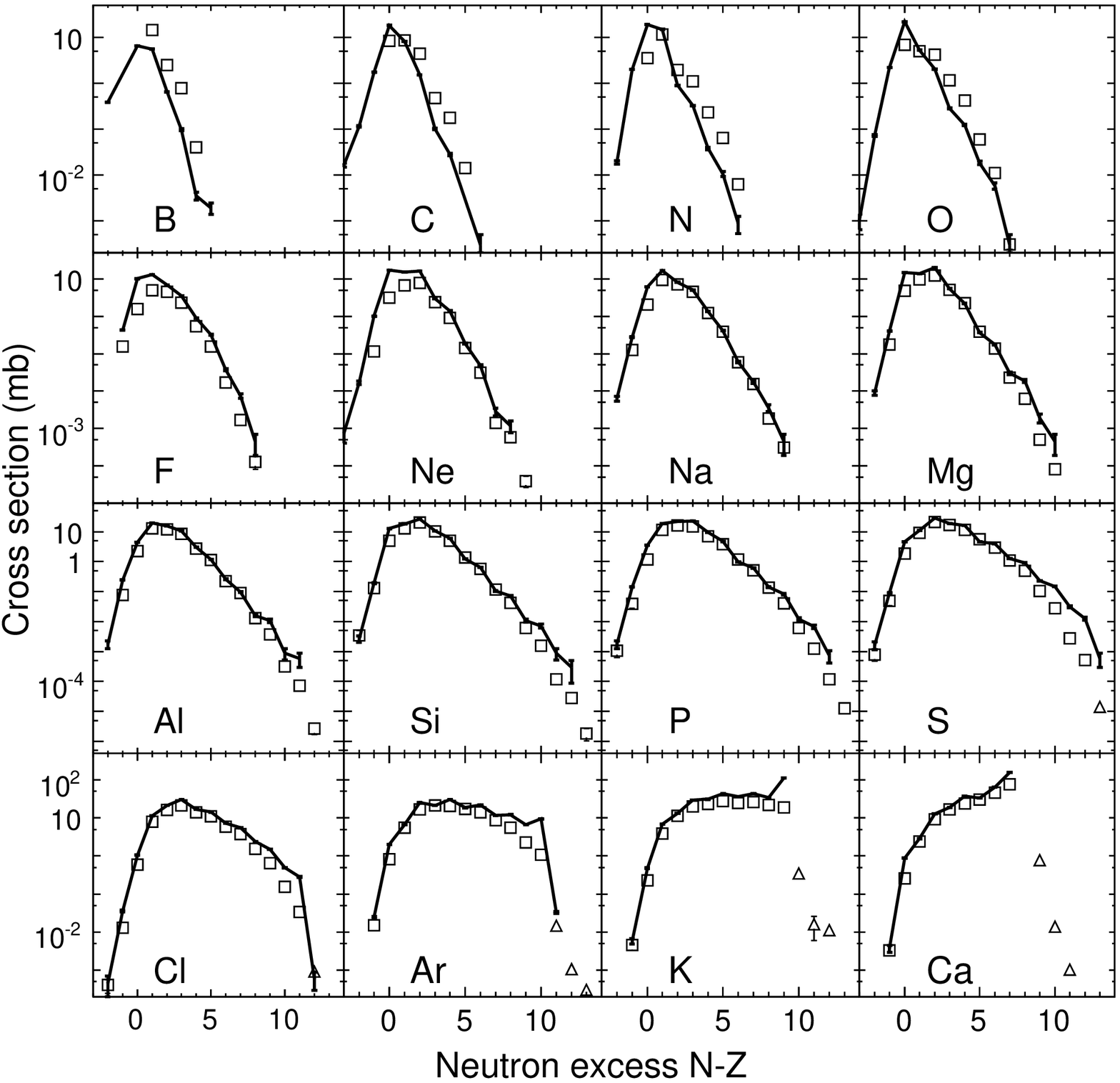} 
\caption{Measured cross sections for $^{48}$Ca fragmentation on $^9$Be
at 140 MeV/nucleon 
\cite{Mocko06,MockoPhD}
compared with LAQGSM03.03 predictions. 
}
\end{figure}
\clearpage            

\begin{figure}[ht]                                                 

\centering
\hspace*{-5mm}
\includegraphics[height=170mm,angle=-0]{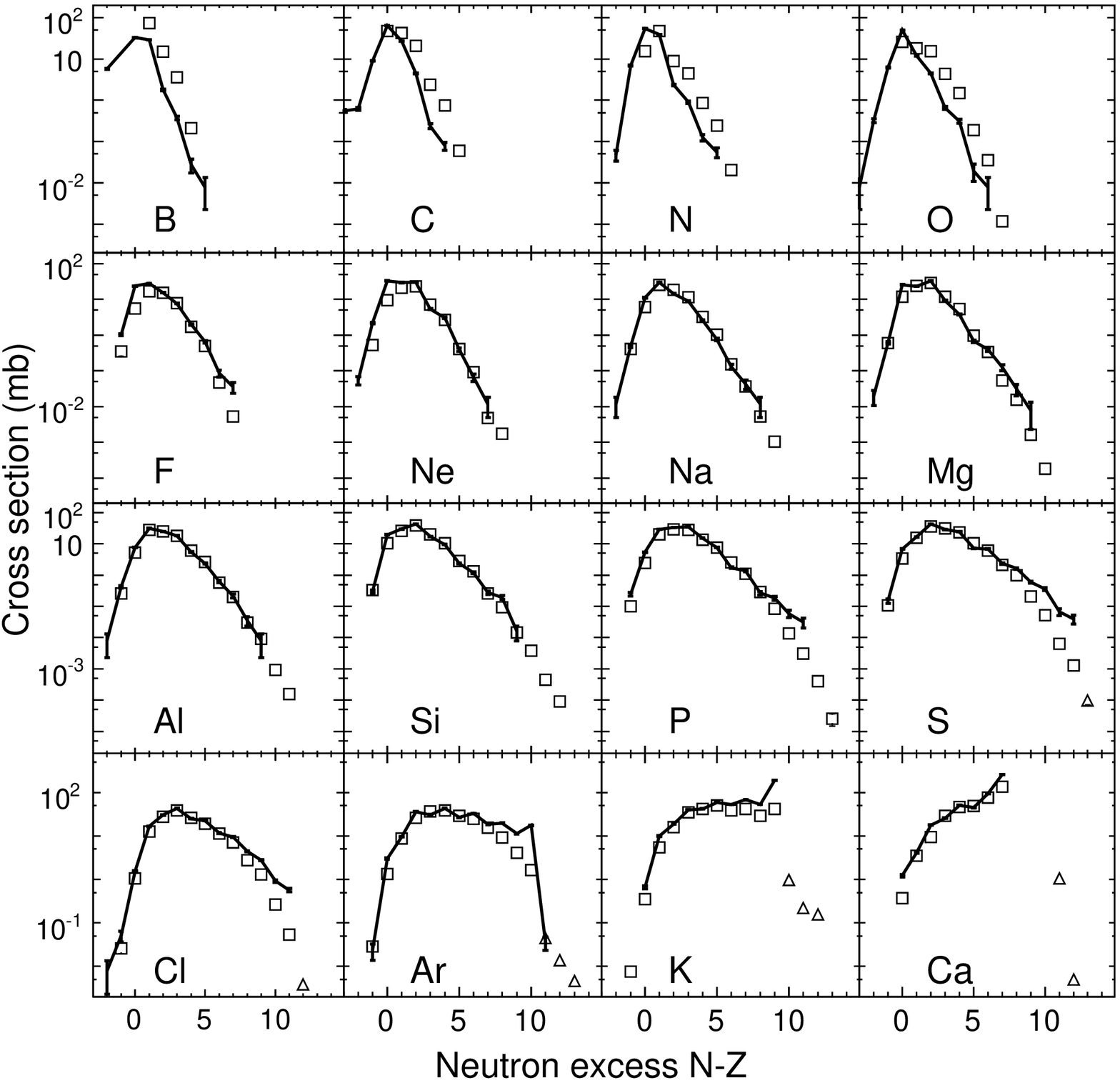} 
\caption{Measured cross sections for $^{48}$Ca fragmentation on $^{181}$Ta
at 140 MeV/nucleon 
\cite{Mocko06,MockoPhD}
compared with LAQGSM03.03 predictions. 
}
\end{figure}
\clearpage            

\begin{figure}[ht]                                                 

\centering
\includegraphics[height=225mm,angle=-0]{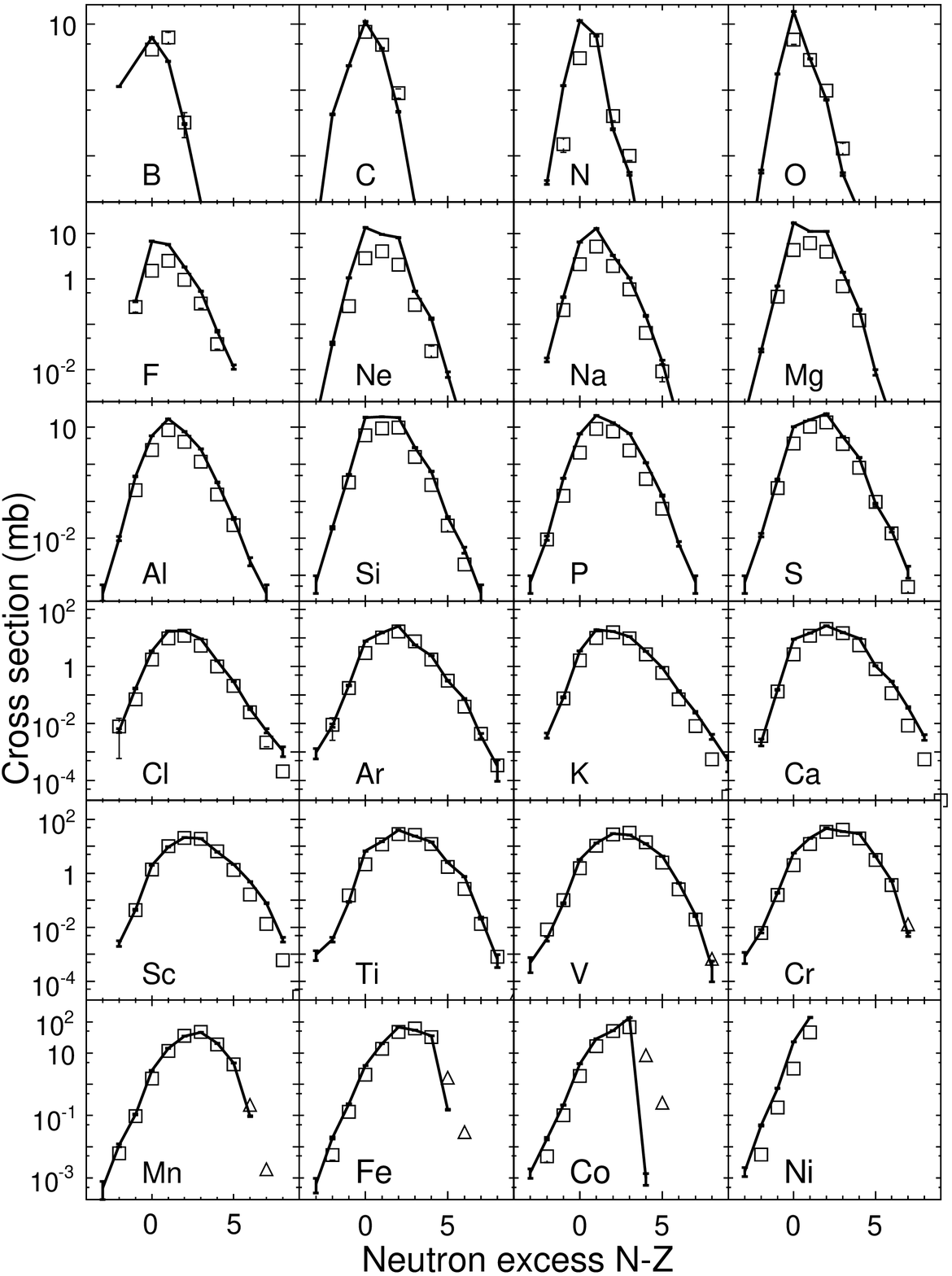} 
\caption{Measured cross sections for $^{58}$Ni fragmentation on $^9$Be
at 140 MeV/nucleon 
\cite{Mocko06,MockoPhD}
compared with LAQGSM03.03 predictions. 
}
\end{figure}
\clearpage            

\begin{figure}[ht]                                                 

\centering
\includegraphics[height=225mm,angle=-0]{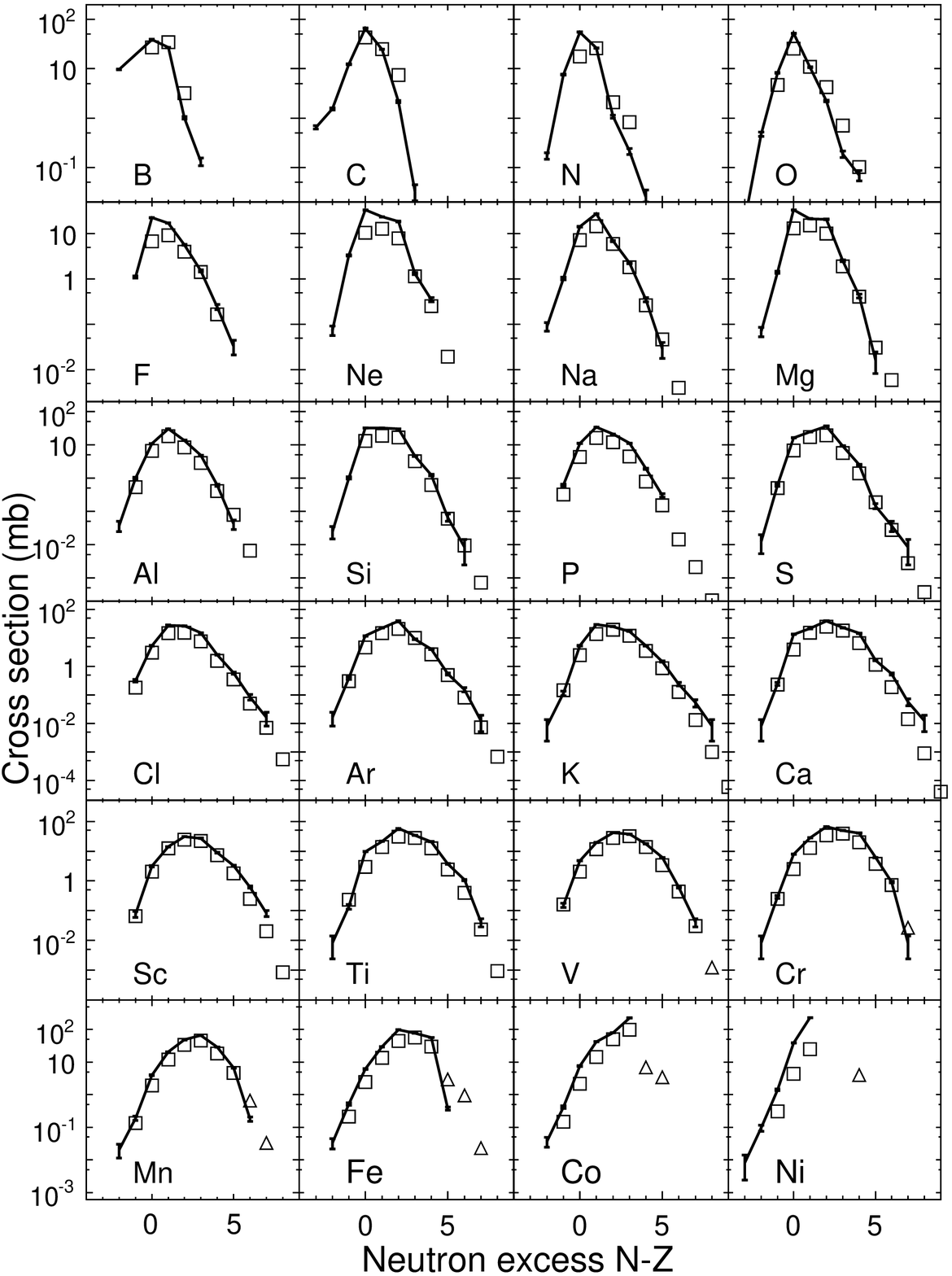} 
\caption{Measured cross sections for $^{58}$Ni fragmentation on $^{181}$Ta
at 140 MeV/nucleon 
\cite{Mocko06,MockoPhD}
compared with LAQGSM03.03 predictions. 
}
\end{figure}
\clearpage            

\begin{figure}[ht]                                                 

\centering
\includegraphics[height=225mm,angle=-0]{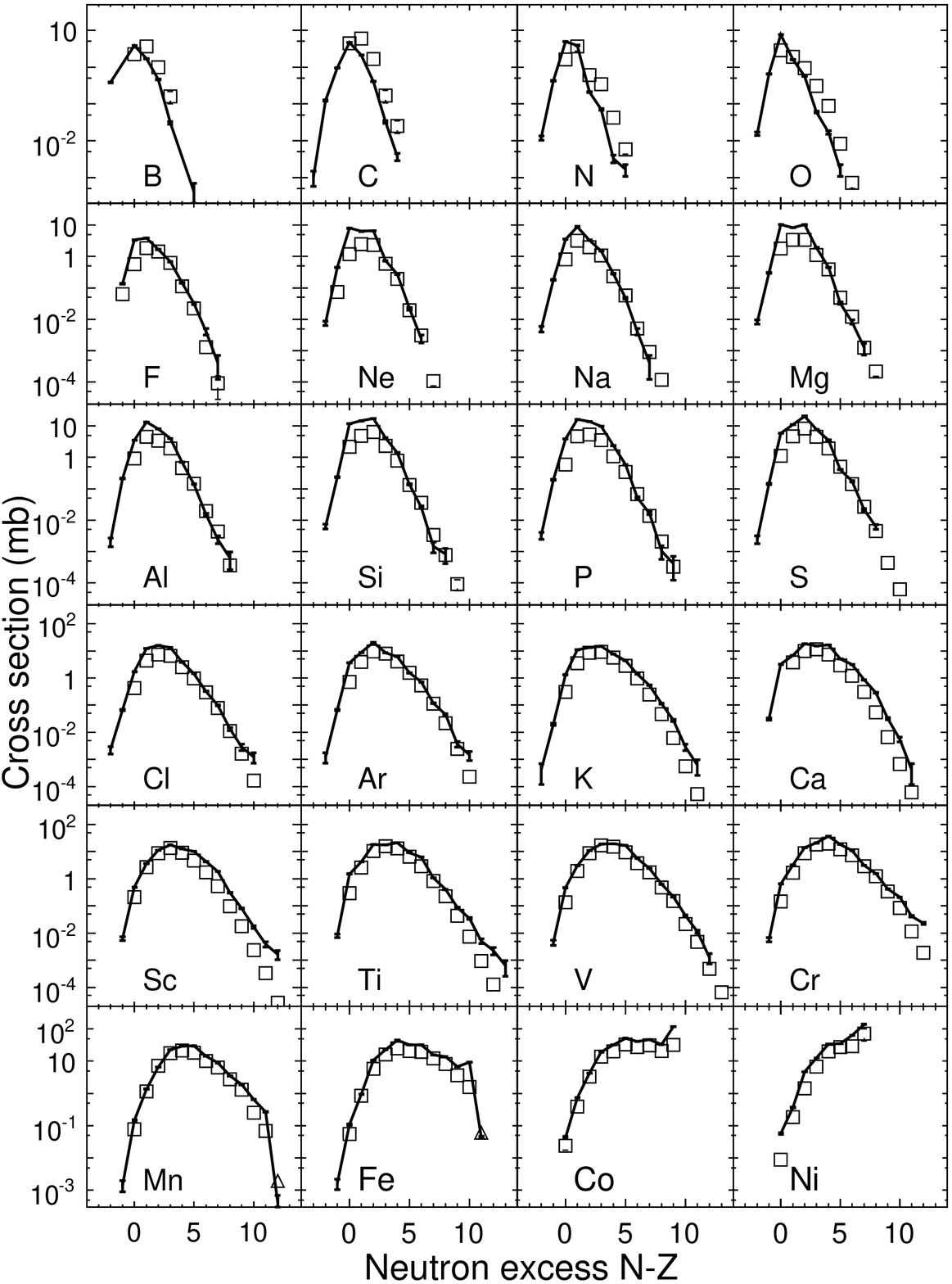} 
\caption{Measured cross sections for $^{64}$Ni fragmentation on $^9$Be
at 140 MeV/nucleon 
\cite{Mocko06,MockoPhD}
compared with LAQGSM03.03 predictions. 
}
\end{figure}
\clearpage            

\begin{figure}[ht]                                                 

\centering
\includegraphics[height=225mm,angle=-0]{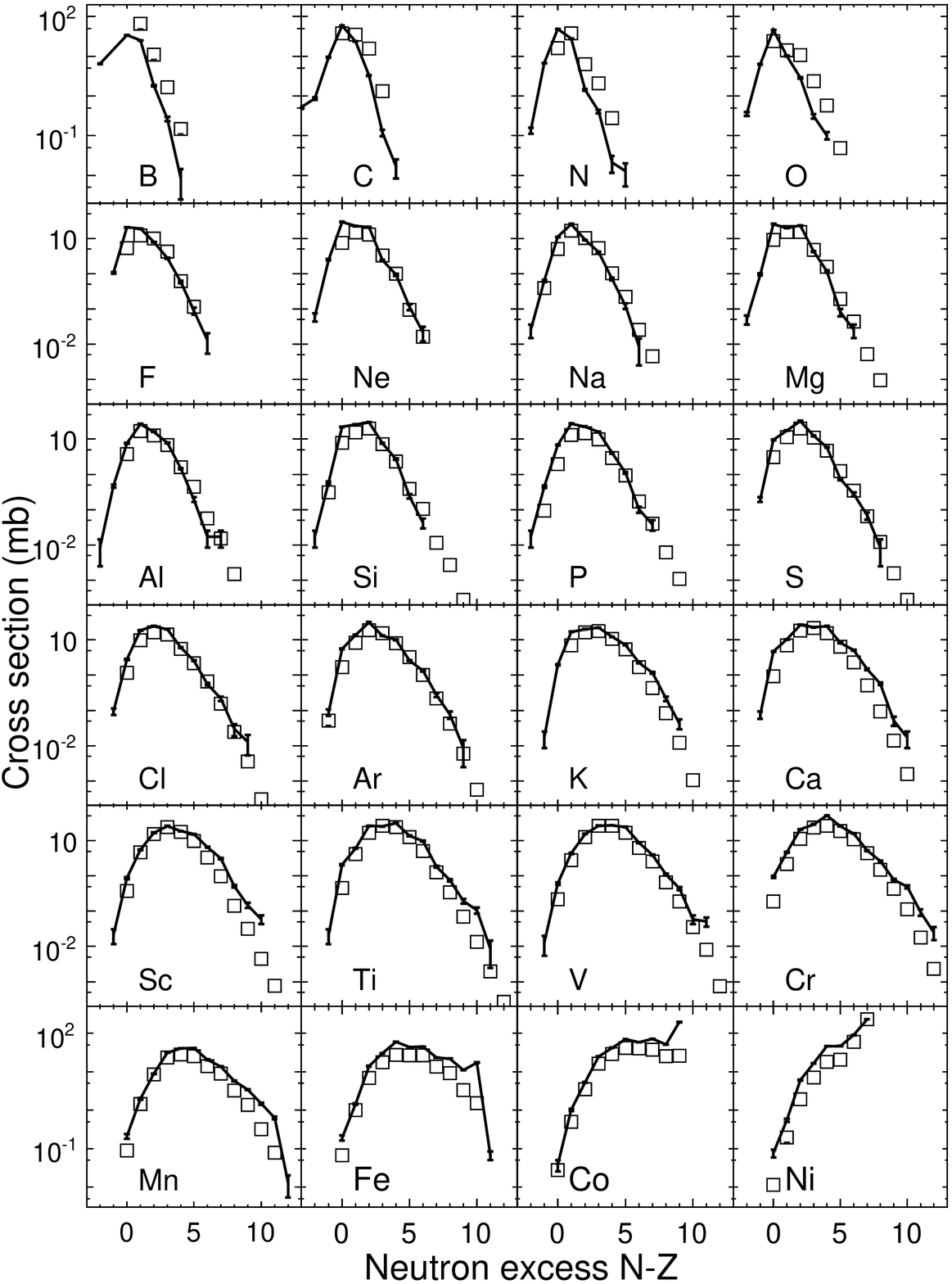} 
\caption{Measured cross sections for $^{64}$Ni fragmentation on $^{181}$Ta
at 140 MeV/nucleon 
\cite{Mocko06,MockoPhD}
compared with LAQGSM03.03 predictions. 
}
\end{figure}
\clearpage            

\begin{figure}[ht]                                                 

\centering
\includegraphics[height=215mm,angle=-0]{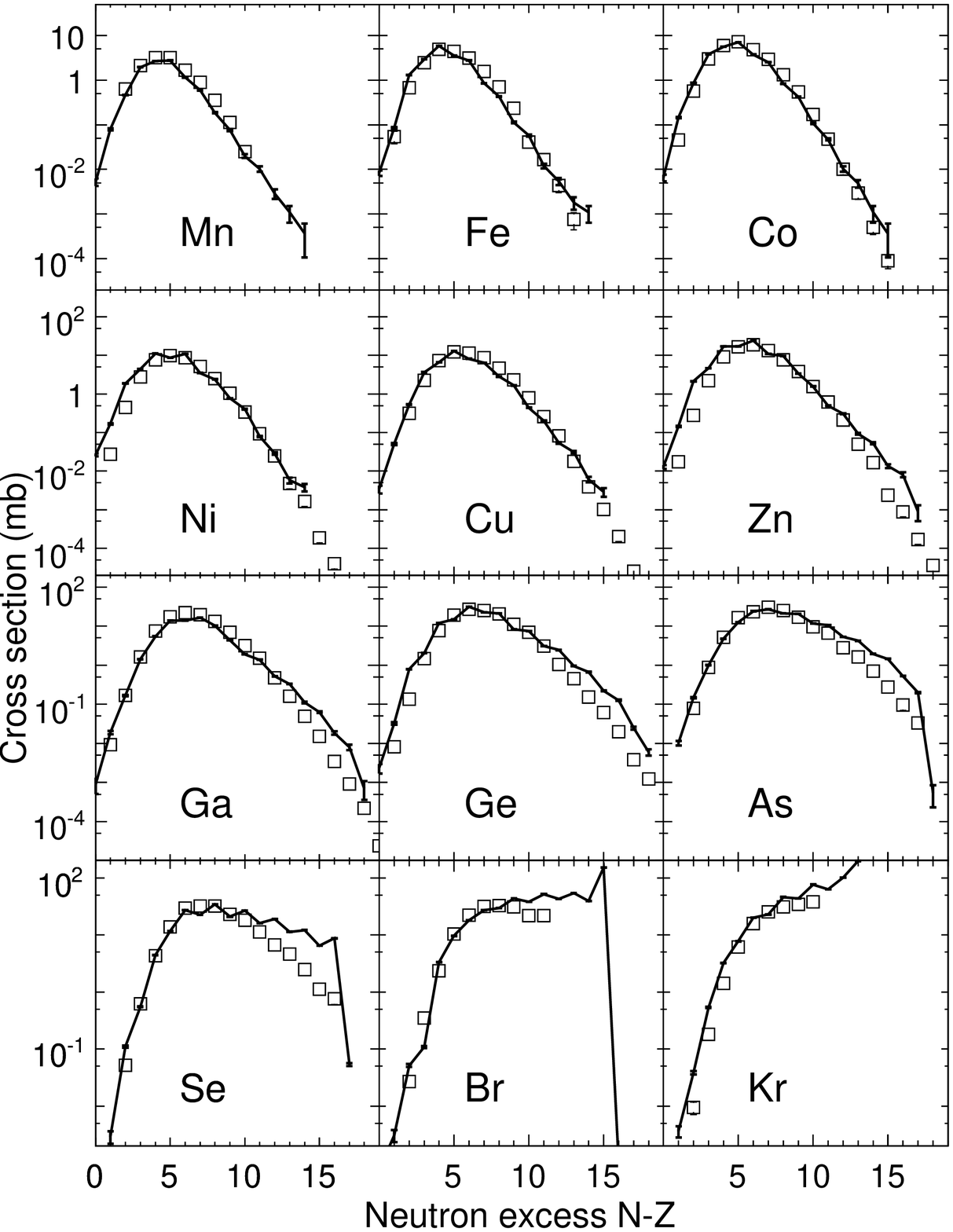} 
\caption{Measured cross sections for $^{86}$Kr fragmentation on $^9$Be
at 64 MeV/nucleon 
\cite{MockoPhD,Mocko07}
compared with LAQGSM03.03 predictions. 
}
\end{figure}
\clearpage            

\begin{figure}[ht]                                                 

\centering
\includegraphics[height=215mm,angle=-0]{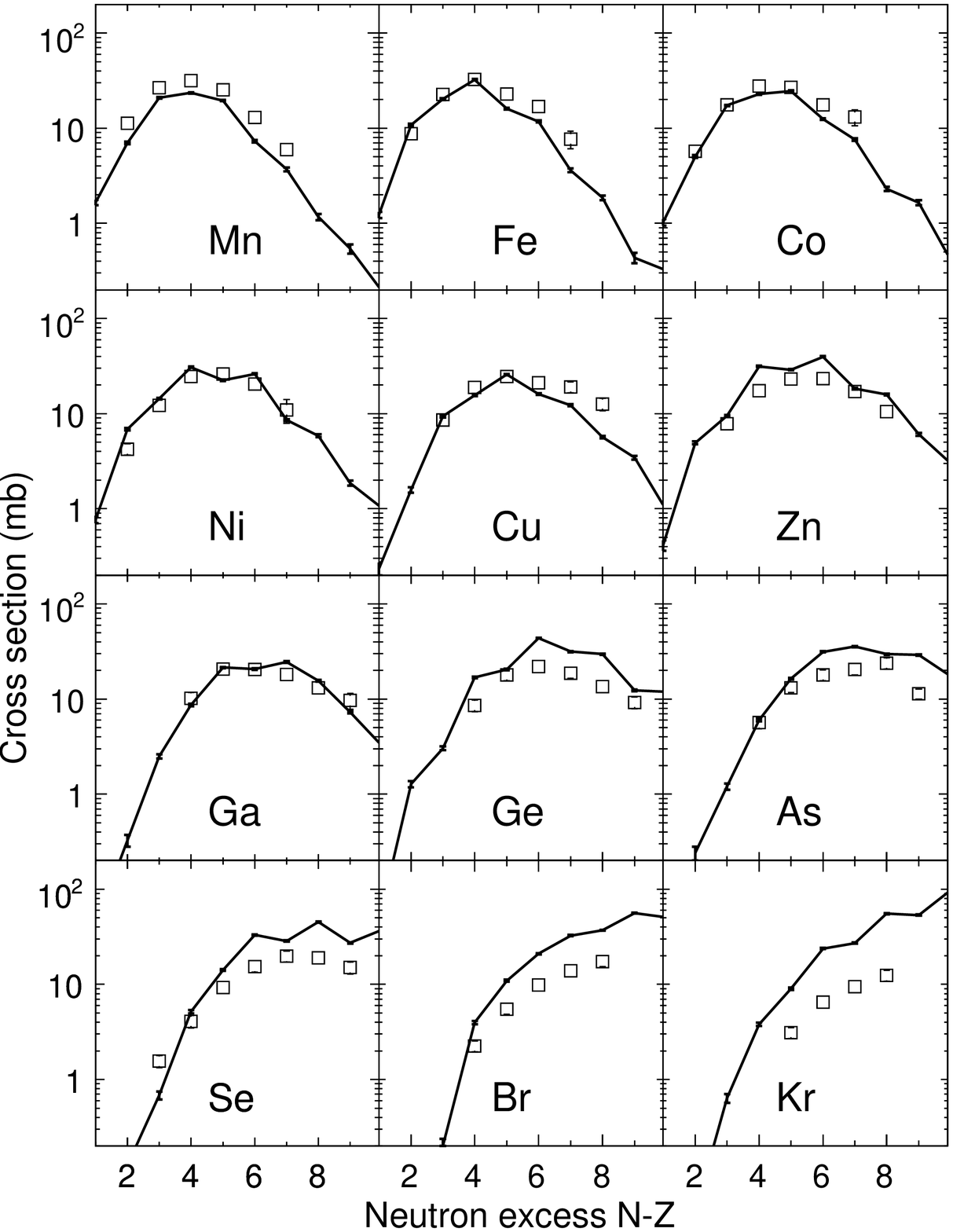} 
\caption{Measured cross sections for $^{86}$Kr fragmentation on $^{181}$Ta
at 64 MeV/nucleon 
\cite{MockoPhD,Mocko07}
compared with LAQGSM03.03 predictions. 
}
\end{figure}

\clearpage            

\end{document}